\documentclass[a4paper,11pt]{article}
\pdfoutput=1 

\usepackage{jheppub} 

\usepackage[T1]{fontenc} 

\usepackage{amsmath,amsfonts,amsbsy,amssymb,array,accents,dsfont,bm,bbm}
\usepackage{enumerate,latexsym,graphicx}
\usepackage{xcolor}

\usepackage{subfigure}
\usepackage{stmaryrd}

\def\cA{\mathcal {A}}  
\def\cD{\mathcal {D}}  
 \def\cH{\mathcal {H}} 
\def\cJ{\mathcal {J}}  \def\cL{\mathcal {L}}
\def\cM{\mathcal {M}}

\newcommand{\eq}[1]{\eqref{#1}}

\newcommand{\cAb}{\ensuremath{\bar{\mathcal{A}}}}

\newcommand{\beq}{\begin{equation}}
\newcommand{\eeq}{\end{equation}}
\newcommand{\bea}{\begin{eqnarray}}
\newcommand{\eea}{\end{eqnarray}}

\newcommand{\vep}{\varepsilon}

\newcommand{\der}{\partial}

\newcommand{\nn}{\nonumber}

\def \sl{\text{sl}}
\def \su{\text{su}}

\newcommand{\n}{{\mathsf n}}
\newcommand{\m}{{\mathsf m}}
\renewcommand{\k}{{\mathsf k}}
\newcommand{\pp}{{\mathsf p}}

\newcommand{\rhoo}{\ensuremath{\! \rho \:\! }}

\usepackage{braket}
\usepackage{tikz}
\usetikzlibrary{matrix,calc,positioning,decorations.markings,decorations.pathmorphing,decorations.pathreplacing}
\usetikzlibrary{arrows,cd}
\usepackage{bbding}
\usetikzlibrary{positioning}
\tikzset{>=stealth}

\newcommand{\sucur}{k} 
\newcommand{\psisl}{\psi}
\newcommand{\psisu}{\chi}

\newcommand{\LLL}{\scriptscriptstyle{\text{L}} }
\newcommand{\RRR}{\scriptscriptstyle{\text{R}} }

\def\tr{{\rm Tr}}

\def\h{\mathfrak{h}}
\def\g{\mathfrak{g}}
\def\u{\mathfrak{u}}
\newcommand{\inv}{^{-1}}
\newcommand{\del}{\partial}             
\newcommand{\bdel}{\bar{\partial}}

\newcommand{\qqquad}{\;, \quad\qquad}  
\DeclareMathOperator{\ch}{ch}
\DeclareMathOperator{\sh}{sh}

\newcommand{\half}{\frac{1}{2}}

\newcommand{\SSS}{\mathbb{S}}

\newcommand{\RR}{\mathbb{R}}
\newcommand{\ZZ}{\mathbb{Z}}

\newcommand{\TT}{\mathbb{T}}
\newcommand{\Aa}{\mathcal{A}}

\newcommand{\Gg}{\mathcal{G}}
\newcommand{\Hh}{\mathcal{H}}

\newcommand{\Jj}{\mathcal{J}}

\newcommand{\Mm}{\mathcal{M}}
\newcommand{\Nn}{\mathcal{N}}

\newcommand{\lL}{\mathrm{l}}
\newcommand{\rR}{\mathrm{r}}

\DeclareMathSymbol{\medhatsym}{\mathord}{largesymbols}{"62} 

\DeclareMathSymbol{\medtildesym}{\mathord}{largesymbols}{"65}

\DeclareMathOperator{\Tr}{Tr}

\newcommand{\R}{\ensuremath{\mathbb{R}}}

\newcommand{\w}{\omega}

\newcommand{\tvphi}{\tilde{\varphi}}
\newcommand{\Z}{\mathbb{Z}}

\usepackage{color}

\usepackage[normalem]{ulem}  

\definecolor{darkred}{rgb}{0.6,0,0}
\definecolor{darkblue}{rgb}{0,0,0.6}

\newcommand{\be}{\begin{equation}}
\newcommand{\ee}{\end{equation}}

\title{\boldmath Black hole microstates from the worldsheet}

\author[a]{Davide Bufalini,}
\author[b,c
]{Sergio Iguri,}
\author[d]{Nicolas Kovensky,}
\author[a]{David Turton}

\affiliation[a]{Mathematical Sciences and STAG Research Centre, University of Southampton, Southampton
SO17 1BJ, United Kingdom.}

\affiliation[b]{Universidad de Buenos Aires, Facultad de Ciencias Exactas y Naturales. Ciudad Universitaria, 1428 Buenos Aires, Argentina.}
\affiliation[c]{CONICET-Universidad de Buenos Aires, Instituto de Astronomía y Física del Espacio (IAFE). C. C. 67, Suc. 28, 1428 Buenos Aires, Argentina.}
\affiliation[d]{Institut de Physique Th\'eorique, Universit\'e Paris Saclay, CEA, CNRS, Orme des Merisiers, 91191 Gif-sur-Yvette CEDEX, France.}

\emailAdd{d.bufalini@soton.ac.uk}
\emailAdd{siguri@iafe.uba.ar}
\emailAdd{nicolas.kovensky@ipht.fr}
\emailAdd{d.j.turton@soton.ac.uk}

\abstract{
Recently an exact worldsheet description of strings propagating in certain black hole microstate geometries was constructed in terms of null-gauged WZW models. In this paper we consider a family of such coset models, in which the currents being gauged are specified by a set of parameters that a priori take arbitrary values. We show that consistency of the spectrum of the worldsheet CFT implies a set of quantisation conditions and parity restrictions on the gauging parameters. We also derive these constraints from an independent geometrical analysis of smoothness, absence of horizons and absence of closed timelike curves. This allows us to prove that the complete set of consistent backgrounds in this class of models is precisely the general family of (NS5-decoupled) non-BPS solutions known as the JMaRT solutions, together with their various (BPS and non-BPS) limits. We clarify several aspects of these backgrounds by expressing their six-dimensional solutions explicitly in terms of five non-negative integers and a single length-scale. Finally we study non-trivial two-charge limits, and exhibit a novel set of non-BPS supergravity solutions describing bound states of NS5 branes carrying momentum charge.
}

\begin{document} 
\maketitle
\flushbottom

\section{Introduction}
\label{sec:intro}

The quantum description of black holes is notably problematic. From the spacetime point of view it remains unclear how to account for their entropy, resolve their singularities, and understand their evaporation.
String theory provides the leading framework within which to address such difficult questions of quantum gravity, and in particular to study black hole microstates. Much progress has been made by constructing supergravity solutions describing these microstates. However, there is rich physics beyond supergravity that may prove crucial to understand the most entropic sector of the black hole Hilbert space.

Bound states of D1 and D5-branes, or of NS5 branes and fundamental strings (F1), possibly also carrying momentum P in a compact direction, have been a very fruitful arena in which to study black hole microstates in string theory. Taking the D1-D5 (or NS5-F1) decoupling limit gives rise to configurations that are asymptotically $AdS_3 \times \SSS^3 \times \cM$, where $\cM$ is $\TT^4$ or K3. This is one of the original examples of holographic duality~\cite{Maldacena:1997re}.

Configurations that have come to be known as circular supertubes~\cite{Balasubramanian:2000rt,Maldacena:2000dr} were important early supergravity solutions describing specific microstates of the two-charge system, in particular in the D1-D5 or NS5-F1 duality frames. These solutions were generalized by Lunin and Mathur, and others, to the full class of two-charge microstates~\cite{Lunin:2001jy,Lunin:2002iz,Taylor:2005db,Kanitscheider:2007wq,Mathur:2018tib}.

An equally important family of
three-charge (D1-D5-P or NS5-F1-P) microstate solutions are known as spectral flowed circular supertubes, of which there are both BPS and non-BPS configurations~\cite{Lunin:2004uu,Giusto:2004id,Giusto:2004ip,Jejjala:2005yu,Giusto:2012yz}. In the $AdS_3$ decoupling limit, the general holographic description of these configurations is well understood \cite{Giusto:2012yz,Chakrabarty:2015foa} and involves spectral flow in the $\mathcal{N}=(4,4)$ superconformal algebra. Moreover, two-charge circular supertubes have proven to be important seed solutions in the construction of much more general families of ``superstratum'' solutions~(see e.g.~\cite{Bena:2015bea,Bena:2016agb,Bena:2016ypk,Bena:2017xbt,Bena:2018bbd,Ceplak:2018pws,Heidmann:2019zws,Giusto:2019qig}).

The non-BPS spectral flowed circular supertubes in the family mentioned above are known as the JMaRT solutions, after the authors of~\cite{Jejjala:2005yu}. These microstates emit ergoregion radiation, which has been interpreted (via holography) as an enhanced, unitary version of Hawking radiation~\cite{Chowdhury:2007jx,Avery:2009tu,Avery:2009xr}. The JMaRT solutions also contain the BPS two-charge circular supertubes and BPS three-charge spectral flowed circular supertubes as (non-trivial) limits. All of these will be included in our analysis. 

Recently a worldsheet description of the JMaRT three-charge NS5-F1-P configurations, in the NS5-brane decoupling limit, was constructed~\cite{Martinec:2017ztd}.
This regime corresponds to little string theory, an example of stringy holography which remains poorly understood.
The models of~\cite{Martinec:2017ztd} make use of a well-known  supersymmetric WZW theory, combining it with the null-gauging formalism. More precisely, they involve an auxiliary (10+2)-dimensional group manifold, which is reduced to the physical (9+1)-dimensional target space by gauging a pair of null chiral currents. 
The corresponding spectrum of perturbative strings and D-branes were studied respectively in~\cite{Martinec:2018nco,Martinec:2019wzw}. The null-gauging construction was further extended to encompass more general Lunin-Mathur solutions~\cite{Martinec:2020gkv}, which correspond to a larger family of gauged sigma models that are generically not cosets. Other coset models that describe wrapped and/or intersecting fivebranes have also recently been studied~\cite{brennan2020wrapped}.

The underlying microscopic configurations involve bound states of NS5 branes (possibly with F1 and/or P charge). Generically, the low-energy supergravity description is not reliable near the fivebrane sources, however the worldsheet theory remains under control. Indeed, these coset theories are exact in $\alpha'$, thus extending the description of these families of black hole microstates beyond the supergravity limit.  

The models considered in~\cite{Martinec:2017ztd,Martinec:2018nco,Martinec:2019wzw} have the following basic structure. The \textit{upstairs} theory (i.e.~before gauging) involves pure NSNS fluxes and is of the form $AdS_3\times \SSS^3\times \R \times \SSS^1\times \TT^4$, where the $\mathbb{R}$ factor is timelike, 
and where the $\SSS^1$ is separated from the $\TT^4$ because it plays a preferred role. Indeed, the gauging does not involve the $\TT^4$ and we shall mostly work in six physical spacetime directions \textit{downstairs} (i.e.~after gauging). The non-trivial WZW model involved is that of the universal cover of $SL(2,\R)$ times $SU(2)$, which constitutes a rich and well studied example of an exactly solvable model \cite{Giveon:1998ns,Kutasov:1999xu,Maldacena:2000hw,Maldacena:2000kv,Maldacena:2001km}. The currents to be gauged are specific null linear combinations of the Cartan currents of $SL(2,\R) \times SU(2)$ and the chiral momenta on  
$\mathbb{R} \times \mathbb{S}^1$, plus a similar (though generically not identical) linear combination of their anti-chiral counterparts.

Within this class of models, it is natural to ask what is the general family of well-behaved backgrounds that can be obtained by considering the most general null linear combination of the currents just described. This question was not addressed in~\cite{Martinec:2017ztd,Martinec:2018nco,Martinec:2019wzw}.
Developing a systematic method for classifying such backgrounds is important for three reasons. First, it offers the possibility of finding novel configurations. Second, it can sharpen our understanding of the general backgrounds and how their parameters are constrained by different consistency conditions, shedding further light on the interplay between worldsheet CFT and spacetime geometry. Third, such techniques may then be applied to other similar classes of gauged models.

In this paper we provide the answer to the above question, by proving that the JMaRT solutions and their limits represent the complete set of supergravity configurations described by this family of coset models. We do so from two complementary but independent points of view, and, in doing so, we clarify several aspects of both the worldsheet models and the supergravity backgrounds. 

First, we consider the most general family of worldsheet coset theories, and derive  necessary and sufficient conditions that lead to a consistent physical spectrum. These consistency conditions are obtained from analysing the gauge orbits, relating different representatives of the same physical operators, combined with worldsheet spectral flow (not to be confused with the spacetime/holographic spectral flow discussed above). This includes not only the spectral flow operation that is an essential part of the $SL(2,\R)$ WZW model \cite{Maldacena:2000hw}, but also that of $SU(2)$. While $SU(2)$ spectral flow does not generate new affine representations, it has proven to be quite useful for string theory applications \cite{Israel:2004ir,Giribet:2007wp,Martinec:2018nco,Martinec:2020gkv}. As it turns out, we obtain constraints that take the form of algebraic relations for the a priori continuous gauging parameters, which imply that they can be written in terms of four integers, $\k,\m,\n,\pp$
(of which only three are independent), plus $R_y$, the continuous modulus corresponding to the asymptotic proper radius of the $\SSS^1$. Furthermore, we derive restrictions on the parities of $\k,\m,\n,\pp$.

We then show that the same conditions can equally be derived from the analysis of the set of supergravity backgrounds obtained from the general gauged models. More precisely, we show that imposing absence of horizons, absence of closed time-like curves (CTCs), and smoothness up to orbifold singularities in the corresponding classical geometries leads to an identical quantisation of the gauging parameters. 

On the other hand, the JMaRT solutions are usually written in terms of their own set of seemingly continuous parameters, which however are known to be constrained by regularity and absence of CTCs to obey their own set of algebraic relations~\cite{Jejjala:2005yu}. This parametrisation is quite awkward to work with, and obscures aspects of the physics.

We find that for the (NS5-decoupled) JMaRT solutions and their limits, one can completely bypass most of the seemingly continuous parameters. Let $n_5$, $n_1$, $n_p$ denote respectively the quantised numbers of NS5 branes, fundamental strings, and units of momentum along $\SSS^1$ present in the background.
We show that, in the NS5-brane decoupling limit, the six-dimensional metric and the NSNS $B$-field can be expressed explicitly in terms of the same set of integers $\k,\m,\n,\pp$ introduced in the coset models, together with $n_5$ and $R_y$. 
Although this result is strongly inspired by our worldsheet analysis, we have derived it independently and purely within supergravity, via a non-trivial manipulation of the above-mentioned algebraic constraints.

For the dilaton, an extra parameter is necessary, which can be taken to be either $n_1/V_4$ or $n_p/V_4$, where $V_4$ is the volume of the $\TT^4$. In the three-charge solutions, there is a constraint that relates $n_1$ and $n_p$, meaning that only one can be chosen independently. We take $V_4$ to be microscopic and fixed, and ignore it when counting parameters, so that we consider the independent ones to be $(\m,\n,n_5,R_y$) plus one of either $\k$ or $\pp$, plus one of either $n_1$ or $n_p$. Without loss of generality, one can restrict the range of the integer parameters to be non-negative.

The resulting expressions for the supergravity fields are identical to those obtained from the general coset worldsheet actions, completing the proof that these are the unique backgrounds that arise in these coset theories. Note that for the latter, $n_5$ defines the level of the $SL(2,\R)$ and $SU(2)$ affine algebras.

The rewriting of these configurations in terms of the integer parametrisation significantly clarifies the properties of these solutions. In particular, it makes some of their symmetries and the action of T-duality manifest. It also sheds light on the somewhat delicate limits that lead to the two-charge configurations where either $n_1$ or $n_p$ is set to zero. As a result, this allows us to derive a novel and non-trivial two-charge non-BPS NS5-P limit of the general solutions in a straightforward way. 

Finally, we also comment on a potential relation to recent investigations of the so-called single-trace $T\bar{T}$ deformation of the (holographic) D1-D5 CFT \cite{Giveon:2017nie,Asrat:2017tzd,Giveon:2017myj}. In the worldsheet model, this can be described by using a null-gauging procedure similar to the  formalism employed throughout this paper, although in that context the $SL(2,\R)$ current involved in the gauging is not in the Cartan subalgebra.   

The structure of this paper is as follows. In Section \ref{sec:review} we introduce the worldsheet models that we consider in this paper, and write down the most general background metric and $B$-field that arise. In Section \ref{sec:CFTanalysis} we review in more detail the $SL(2,\R)$ and $SU(2)$ WZW models, and analyse the consistency of the CFT spectrum in terms of the gauging parameters. 
In Section \ref{sec:gen-CTC} we analyse absence of horizons, absence of CTCs, and smoothness in the corresponding supergravity backgrounds. In Section \ref{sec:JMaRTuniqueness} we firstly match the resulting worldsheet models to the general JMaRT solutions. We then discuss in detail their various limits, including two-charge (non-BPS), BPS, and $AdS_3$ limits. In Section \ref{sec:discussion} we further discuss our results.

\section{A class of null-gauged WZW models}
\label{sec:review}

In this section we review some relevant aspects of the models that we shall study in this work. We will aim to be brief where possible; the interested reader can find the details in the works~\cite{Martinec:2017ztd,Martinec:2018nco,Martinec:2019wzw,Martinec:2020gkv}.

As discussed in the Introduction, we consider the (10+2)-dimensional upstairs target $AdS_3\times \SSS^3\times \R_t \times \SSS^1_y \times \TT^4$, where we have introduced coordinates $t$ and $y$ for the timelike $\R$ and spacelike $\SSS^1$ factors respectively.
Since the Cartan direction in $SL(2,\R)$ is timelike, and that of $SU(2)$ is spacelike, and the levels are the same, one can form null linear combinations $J^3_\sl \pm J^3_\su$ in both holomorphic and antiholomorphic sectors of the worldsheet theory.  Gauging such null currents leads~\cite{Israel:2004ir} to the background sourced by a circular array of NS5 branes on their Coulomb branch~\cite{Sfetsos:1998xd,Giveon:1999px,Giveon:1999tq}.

The (10+2)-dimensional models have other null currents that are linear combinations of $J^3_\sl$, $J^3_\su$, $\partial_t$ and $\partial_y$. It was recently found that particular linear combinations of these currents give rise to a family of backgrounds that include NS5-P and NS5-F1 BPS circular supertubes~\cite{Maldacena:2000dr,Balasubramanian:2000rt}, as well as NS5-F1-P BPS and non-BPS spectral flowed supertubes~\cite{Giusto:2004id,Giusto:2004ip,Jejjala:2005yu,Giusto:2012yz,Chakrabarty:2015foa}. 

We now review the basics of these constructions and then consider the most general pair of such chiral null currents that can be gauged. We derive the corresponding supergravity fields, written as functions of the general gauging parameters. 

\subsection{Null-gauged sigma models}
\label{sec:nullgauging}

We now briefly review the null gauging formalism for general sigma models, before specialising to WZW models (see e.g.~\cite{Tseytlin:1993my,Klimcik:1994wp,Hull:1989jk,Figueroa-OFarrill:2005vws}). In this passage we follow the presentation of~\cite{Martinec:2020gkv}. We use units in which $\alpha'=1$, and  work at tree level in the string coupling $g_s$.


Consider the string worldsheet $\Mm_2$ and an embedding map $\varphi $ into a pseudo-Riemannian manifold $\Nn$, namely $\varphi: \Mm_2 \to \varphi(\Mm_2) \subset \Nn$. The target manifold $\Nn$ is endowed with a metric with components $G_{ij}$. We wish to gauge a set of Killing vectors $\xi_a$ generating isometries of $\Nn$, where $a$ labels the different Killing vectors (in this paper we will have $a=1,2$). We introduce a set of independent worldsheet gauge fields $\cA^a$, one corresponding to each Killing vector. 
Then the kinetic term in the string sigma model action is written in terms of the covariant derivative
\be \label{eq:cov-der}
\cD\varphi^i = \partial\varphi^i - \cA^a \xi_a^i
\ee
and takes the form
\be
\label{gauged KE}
\cL_{\rm K} = \mathcal{D} \varphi^i \, G_{ij} \, \overline{\mathcal{D}} \varphi^j
\;=\; (\partial \varphi^i - \cA^a \:\! \xi_a^i ) \, G_{ij} \,  (\bar\partial \varphi^j - \cAb^a \:\! \xi_a^j) \;.
\ee
To write the gauged Wess-Zumino (WZ) term we introduce target-space one-forms $\theta_a$ (we follow the notation of~\cite{Figueroa-OFarrill:2005vws}), pulled back to the worldsheet. The WZ term can then be written as 
\be
\label{eq:g-wz-term}
\cL_{\rm WZ} = B_{ij} \partial \varphi^i \bar\partial\varphi^j + \cA^a \theta_{a,i} \bar\partial\varphi^i - \cAb^a \theta_{a,i} \partial\varphi^i
+ \xi_{[a}^i \theta^{~}_{b],i} \cA^a \cAb^b \;
\ee
where $\theta_{a,i}$ denotes the $i^{\text{th}}$ component of the one-form $\theta_a$. For our null-gauged models, the target-space one-forms $\theta_a$ are given by
\be \label{eq:theta-xi}
\qquad\quad    \theta_{a} \;=\; (-1)^{a+1}  \xi_{a} \cdot d\varphi  \;\equiv\; (-1)^{a+1} \xi_a^i G_{ij} d\varphi^j \,,\:\!      \qquad\quad (a=1,2)\,.
\ee
For a consistent gauging, the following conditions must hold:
\be
\imath_a H \;=\;  d\theta_a \,,\qquad \imath_a\theta_b \;=\; - \imath_b \theta_a   \,,
\ee
where $H = dB$. The expression \eq{eq:theta-xi} implies that half of the gauge field components decouple, so that they are naturally chiral. 
As a result, in our $U(1)\times U(1)$ gauged models, the coefficient of the term quadratic in gauge fields is proportional to the quantity 
\be
\Sigma \;\equiv\; -\frac12\xi_1^i G_{ij} \xi_2^j   \,.
\label{eq:sigma-def}
\ee 
All together, the terms in the action involving the gauge fields then reduce to 
\be \label{eq:g-terms-0}
\cL_{\cA} \;=\; -2\cA^2 \xi_2^i G_{ij} \bar\partial\varphi^j -2 \cAb^1 \xi_1^i G_{ij} \partial\varphi^j - 4 \cA^2 \cAb^1 \Sigma  ~,
\ee
and in the following, we shall denote $\cA\equiv\cA^2$, $\cAb\equiv\cAb^1$. 

We define the worldsheet currents $\cJ$, $\bar{\cJ}$ to be pull-backs of the target-space one-forms as follows\footnote{Note that $\cJ,\bar\cJ$ are related to the CFT current operators $J, \bar{J}$ by factors of $i$, e.g.~$J \, =\, i{\cal{J}}$, see Eq.~\eqref{Jcurrents}.}
\be
\label{eq:J-def}
\cJ \,\equiv\, - \theta_{1} \cdot \partial\varphi \,\equiv\, - \theta_{1,i} \:\! \partial\varphi^i \;, \qquad \bar\cJ \,\equiv\, \theta_{2} \cdot \bar\partial\varphi \,\equiv\, \theta_{2,i} \:\!\bar\partial\varphi^i \,.
\ee
By using \eq{eq:theta-xi}, one can then write the gauge terms \eq{eq:g-terms-0} as
\begin{align}
\label{eq:g-terms-gen}
\cL_{\cA} \;=\; 2\cA \;\! \theta_{2,i} \:\!\bar\partial\varphi^i -2 \cAb \;\! \theta_{1,i} \:\! \partial\varphi^i -4 \cA \cAb \:\! \Sigma 
~\,\equiv\,~ 2 \cA \bar \cJ + 2 \bar\cA \cJ - 4 \cA \bar\cA \Sigma\,.
\end{align}
Upon integrating out the gauge fields, the gauge terms in the action then become
\be \label{eq:g-terms-int-out}
\frac{\cJ \bar\cJ}{\Sigma} ~=~ - \frac{1}{\Sigma} \big( \theta_{1} \cdot \partial\varphi  \big) \big( \theta_{2} \cdot \bar\partial\varphi \big) 
~=~ \frac{1}{\Sigma} \big( \xi_{1} \cdot \partial\varphi  \big) \big( \xi_{2} \cdot \bar\partial\varphi \big) \,,
\ee
where $\xi_{1} \cdot \partial\varphi \equiv\xi_1^i G_{ij} \partial\varphi^j$.
Thus the overall effect of the null gauging procedure is to add the term \eq{eq:g-terms-int-out} to the ungauged sigma model lagrangian.

\subsection{Null-gauged WZW models}

We now specialise the discussion to the case where the upstairs theory is a WZW model whose target space is a Lie group $\Gg $, and thus we replace $\varphi$ with a $\Gg$-valued function $g:\Mm_2 \to \Gg$. We will shortly consider $\Gg $ to be a direct product of simple and abelian factors, but for the moment we focus on one of the simple factors. We follow in places the presentation in~\cite{Martinec:2019wzw}.

We wish to gauge the action of a subgroup $ \Hh \subset \Gg$. 
Its action on $\Gg$ is defined by the group homomorphism embeddings 
\begin{equation}
\ell : \Hh \to \Gg_L \qqquad r : \Hh \to \Gg_R \ ,
\end{equation}
where $ \Gg_L \times \Gg_R $ is the standard left-right isometry group, and such that we will gauge the transformations 
\be
\label{eq:Haction}
g ~\mapsto~ \ell(h) \,g\; r(h)^{-1} \,, \qquad h\in \cH \,.
\ee
The group embeddings $\ell$ and $r$ induce corresponding Lie algebra homomorphisms. Since the meaning will be clear from the context, it is convenient to abuse notation and re-use the same symbols $\ell$ and $r$ for the induced Lie algebra homomorphisms,
\begin{equation}
\ell : ~\h \to \g \qqquad r : ~\h \to \g \,.
\end{equation}
To write the corresponding Killing vector field, let us denote the left(right)-invariant vector field corresponding to a generic $X\in\mathfrak{g}$ by $X^L$ ($X^R$). Given a basis of $\h$, for each element $X_a$ there is a corresponding Killing vector field given by (see e.g.~\cite{Figueroa-OFarrill:2005vws})
\be
\label{Killingvec}
\xi_a ~\equiv~ - \ell(X_a)^R-  r(X_a)^L   \,.
\ee
Let us write the left and right Maurer-Cartan one-forms as
\begin{equation}
\theta_L = g\inv dg\qqquad \theta_R = - \, dg \, g\inv \,.
\end{equation}
We denote by
$ \braket{\cdot,\cdot} $ the standard inner product on $\g$  given by the Killing form. More explicitly, for matrix groups we use the normalisation $\langle A,B\rangle = \tr(AB)$.
In terms of these, the one-forms $\theta_a$ introduced in Eqs.\;\eq{eq:g-wz-term}--\eq{eq:theta-xi} take the form \begin{equation}
\theta_a = \braket{\ell(X_a), \theta_R} - \braket{r(X_a),\theta_L}\,.
\end{equation}

\subsection{The models we study}

We work in Type II superstring theory, however we suppress worldsheet fermions in this section for ease of presentation.
Worldsheet fermions will be discussed in detail in Section \ref{sec:CFTanalysis} below. We consider the cosets\footnote{More precisely, the upstairs model involves the universal cover of $SL(2,R)$, and globally we gauge
$\mathbb{R} \times U(1)$, as we discuss in more detail in Section~\ref{sec:consistency-CFT}; see also~\cite{Martinec:2018nco}.}
\begin{equation}
\label{eq: coset}
\Gg/\Hh \; \times \TT^4 = \frac{SL(2,\R) \times SU(2)  \times \R_t \times U(1)_y}{U(1)_L \times U(1)_R} \; \times \TT^4 \,.
\end{equation}
To define the action of $\Hh = U(1)_L \times U(1)_R$ we must specify the embedding into each of the four subgroups of the upstairs group $\Gg$ that participate in the gauging. Parametrising $SL(2,\R)$ as $SU(1,1)$, we introduce coordinates for the upstairs subgroup elements as
\begin{align}
\label{eq: coordinates group element}
    g & \;=\; \big( g_\sl, \; g_\su, \; g_t, \; g_y \big)  \nonumber\\
    & \;=\; \left( e^{\frac{i}{2}(\tau - \sigma) \sigma_3 } e^{\rho \sigma_1} e^{\frac{i}{2} (\tau + \sigma) \sigma_3}, \;
    e^{\frac{i}{2}(\psi - \phi) \sigma_3 } e^{i \theta \sigma_1} e^{\frac{i}{2} (\psi + \phi) \sigma_3}, \;
    e^{t}, \;
    e^{\frac{i\:\! y}{R_y}}  \right) \, ,
\end{align}
where $\sigma_i$ denotes the $i^{\text{th}}$ Pauli matrix and $ y \in [0,2\pi R_y)$. 
At the level of the algebra, the chiral embeddings we consider are specified by eight arbitrary real parameters $\lL_i \,, \rR_i$, $i=1,2,3,4$, as follows (the ordering of subgroups is as in Eq.\;\eqref{eq: coordinates group element}),
\begin{gather}\label{eq:gauge-action-2}
\begin{aligned}
\ell(\alpha) &\;=\;~\Big(i \, \mathrm{l}_1 \alpha  \sigma_3,\; -i \, \lL_2 \alpha  \sigma_3 ,\;  \, \lL_3  \alpha,\; - i \frac{\lL_4 }{R_y} \, \alpha \Big)
\qqquad r(\alpha) \,=\, 0 \,, \cr
r(\beta) &\;=\; -\Big(i\, \rR_1  \beta \sigma_3,\;  -i \, \rR_2  \beta \sigma_3 ,\;  \, \mathrm{r}_3 \beta,\; - i\frac{ \rR_4 }{R_y} \, \beta   \Big) \qqquad
\ell(\beta) \,=\, 0 \,,
\end{aligned}
\end{gather}
\noindent where $\alpha, \beta \in \R$ and the signs have been chosen for later convenience, in particular for Eq.\;\eqref{eq:jjbar-def} below.
The group action \eqref{eq:Haction} being gauged is then 
\begin{equation}
g  \;\mapsto\; \left(e^{{i} \lL_1 \alpha\sigma_3} \:\! g_\sl \, e^{{i} \rR_1 \beta \sigma_3} \,,\,
{ e^{-{i} \lL_2 \alpha \sigma_3} \:\! g_\su \, e^{-{i} \rR_2 \beta\sigma_3}} \, , \,
 e^{\lL_3 \alpha} g_t \,
e^{\rR_3 \beta} \,,\,
 e^{-i \frac{\lL_4}{R_y} \alpha} 
 g_y \,
e^{-i \frac{\rR_4}{R_y} \beta} \right) \; .
\end{equation}
The general gauge-invariant action for such asymmetric cosets can be found in~\cite{Quella:2002fk}. We introduce two (independent) $\h$-valued worldsheet gauge fields  $(\cA_1,\cAb_1)$ and $(\cA_2,\cAb_2)$.
The gauged WZW action takes the form
\begin{gather}
S \;=\; \sum_j\text{sgn}(\kappa_j) \frac{k_j}{\pi}\Bigg( \int\limits_{~\Mm_2} \half \Tr\left[g\inv \del g g\inv \bdel g \right]_j d^2z + i\, \int\limits_{~\Mm_3} \frac{1}{3!} \Tr\left[ g\inv dg \:\! \wedge g\inv dg \wedge g\inv dg   \right]_j     \nonumber\\
\qquad {} + \!\! \int\limits_{~\Mm_2} \! \Tr \left[ - \sum\limits_{a=1}^2 \left[ \ell(\cAb_a) \del g g\inv \right]_j +  \sum\limits_{a=1}^2 \left[ r(\cA_a) g\inv \bdel g \right]_j - \sum\limits_{a,b=1}^2 \left[ g\inv \ell(\cAb_a) g \;\! r(\cA_b) \right]_j
\right] d^2 z \!\!\: \Bigg) ,
\end{gather}
where $j$ runs over the Lie algebras, $\Mm_3$ is a three-dimensional auxiliary space such that $\Mm_2 = \del \Mm_3$, $k_j$ are the levels of the Kac-Moody algebras, and $\text{sgn}(\kappa_j)$ are the signatures of the respective Killing forms, which in our conventions is positive for $SL(2,\RR)$ and negative for ($SU(2)$, $\RR_t$, $\SSS^1_y$). Here the embeddings $\ell$, $r$ should be understood as corresponding to each respective Lie subalgebra, i.e.~the components of the right-hand sides of Eq.\;\eqref{eq:gauge-action-2}.

We now recall from the discussion of general gaugings of sigma models in Eqs.\;\eq{eq:cov-der}--\eq{eq:g-terms-int-out} that, since the gauge fields are null and chiral, one of their components simply drops out, such that we can set
\be
\cA_1=0 \qqquad \cAb_2=0\,.
\ee
The gauge field embeddings are then
\begin{gather}\label{eq:gauge-action}
\begin{aligned}
\ell(\cAb_1) &\;=\; ~ \Big(i \, \lL_1  \bar{\Aa}_1 \sigma_3,\; -i \, \lL_2  \bar{\Aa}_1 \sigma_3 ,\;  \, \lL_3 \, \bar{\Aa}_1,\; - i \frac{\lL_4}{R_y}\,  \bar{\Aa}_1 \Big)
\qqquad \ell(\cA_2) \,=\, 0 \,, \cr
r(\cA_2) &\;=\; -\Big(i \, \rR_1  \cA_2 \sigma_3,\; - i \, \rR_2 \cA_2 \sigma_3 ,\;  \rR_3 \cA_2,\; -i\frac{ \rR_4}{R_y}\, \cA_2 \Big)
\qqquad r(\cAb_1) \,=\, 0 \,, 
\end{aligned}
\end{gather}
consistently with \eqref{eq:gauge-action-2}. As before, in order to lighten the notation we set $\cA=\cA_2$, $\cAb=\cAb_1$ from now on.

We introduce the currents (our conventions follow \cite[App.~A]{Martinec:2020gkv})
\begin{align}
	\, \mathsf{j}^{3}_{\sl} & = k_{\sl} \Tr \left( -i\frac{\sigma_3}{2} \,  \del g_{\sl} \, g\inv_{\sl}  \right) \qqquad
	\bar{\, \mathsf{j}}_{\sl}^3  = k_{\sl} \Tr \left(  -i\frac{\sigma_3}{2} \, g\inv_{\sl}  \,  \bdel g_{\sl}  \right) \,,
\end{align}
and similarly for $SU(2)$. Their explicit form in our coordinates is
\begin{align}
\begin{aligned}
    & \, \mathsf{j}^3_\sl  = n_5 \big( \cosh^2\rho \; \del \tau + \sinh^2 \rho \;  \del \sigma  \big) \qqquad \bar{\, \mathsf{j}}^3_\sl = n_5 \big( \cosh^2\rho \;  \bdel \tau - \sinh^2 \rho \;  \bdel \sigma  \big)\, , \\
    & \mathsf{j}^3_\su = n_5 \big(  \cos^2 \theta \; \del \psi - \sin^2 \theta\;  \del \phi \big) \qqquad \quad \bar{\, \mathsf{j}}^3_\su = n_5 \big(   \cos^2 \theta \; \bdel \psi + \sin^2 \theta \; \bdel \phi \big) \, .
\end{aligned}
\end{align}
We also define
\begin{align}
	\mathsf{P}^t_{L}= \del t \ , \qquad \mathsf{P}^t_{R} = \bdel t \ , \qquad
	 \mathsf{P}^y_{L}= \del y \ , \qquad \mathsf{P}^y_{R}= \bdel y \,.
\end{align}
Note that, as usual, the bosonic subsector of the supersymmetric WZW model has $k_{\sl}=n_5 - 2$ and $k_{\su} =n_5 + 2$ while the full supersymmetric model has $k_{\sl}=k_\su=n_5$. As noted above, we are suppressing worldsheet fermions in the present section.
The shift in the levels is important (see e.g.~the discussion in~\cite{Martinec:2020gkv}), and we will take care of this in detail when discussing results in the worldsheet CFT in the next section. When discussing supergravity solutions we will work in the usual supergravity regime  $n_5 \gg 1$ (and in the fivebrane decoupling limit $g_s \to 0$) and thus for our purposes in this section we can simply work with $k_{\sl}=k_\su=n_5$. To have canonical kinetic terms we set the (otherwise irrelevant) $\widehat{\u(1)}$ levels to be $k_t = 2, \,  k_y = 2R_y^2\;\!$.

The group action that we gauge, defined in Eq.\;\eqref{eq:gauge-action-2}, corresponds to gauging the currents
\begin{align}
\begin{aligned}
\label{eq:jjbar-def}
	\Jj & = \lL_1 \, \mathsf{j}^3_\sl +\lL_2 \, \mathsf{j}^3_\su + \lL_3 \mathsf{P}^t_{\LLL} +\lL_4 \mathsf{P}^y_{\LLL}  \, , \cr
	\bar{\Jj} & = \rR_1 \bar{\, \mathsf{j}}^3_\sl +\rR_2 \bar{\, \mathsf{j}}^3_\su + \rR_3 \mathsf{P}^t_{\RRR} +\rR_4 \mathsf{P}^y_{\RRR} \,,
\end{aligned}
\end{align}
which we require to be null by imposing
\begin{equation}\label{eq: null gauge constraints}
 n_5 (\mathrm{l}_1^2 - \lL_2^2) + \lL_3^2 - \lL_4^2 = 0 \qqquad 
 n_5 (\rR_1^2 - \rR_2^2) + \rR_3^2 - \rR_4^2 = 0 \,.
\end{equation}
One can  use these constraints to fix the overall normalization of the gauging parameters. 
We assume that $\lL_1=\rR_1 \neq 0$ and divide through by $\lL_1^2$ and $\rR_1^2$, to work with the ratios 
\begin{equation}\label{eq: null gauge constraints-2}
 l_i \;=\; \frac{\lL_i}{\lL_1} \,, 
 \qquad \quad r_i \;=\; \frac{\rR_i}{\rR_1} \,, \qquad i=2,3,4\,.
\end{equation}
In practice this has the same effect as setting $\lL_1 = \rR_1 = 1$, however we have introduced a separate notation for later convenience. Of course, one can modify this step accordingly to deal with models in which $\lL_1=\rR_1=0$.
For later use we record that the ratio parameters $l_i$, $r_i$, $i=2,3,4$ are subject to the constraints
\begin{equation}\label{eq: null gauge constraints-3}
 n_5 (1 - l_2^2) + l_3^2 - l_4^2 = 0 \qqquad 
 n_5 (1 - r_2^2) + r_3^2 - r_4^2 = 0 \,.
\end{equation}

\medskip

In the upstairs model, the line element and NSNS three-form flux are given by 
\begin{align}
    ds^2 & = n_5 \big( - \cosh^2 \rho d\tau^2 + d\rho^2 + \sinh^2 \rho d\sigma^2       + d\theta^2 + \cos^2 \theta d\psi^2 + \sin^2 \theta d\phi^2   \big) - dt^2 + dy^2 , \nonumber\\
    H & = n_5 \big( \sinh2\rho \; d\rho \wedge d\tau \wedge d\sigma 
    %
    +
    \sin2\theta \;  d\theta \wedge d\psi \wedge d\phi \big).
\end{align}
The Killing vectors associated to the group action \eqref{eq:Haction} being gauged are
\begin{align}
\begin{aligned}
\label{eq: killing vectors}
	\xi_{\LLL} & =  (\del_\tau - \del_\sigma) - l_2(\del_\psi - \del_\phi) + l_3 \del_t - l_4 \del_y \,,\cr
	\xi_{\RRR} & =  (\del_\tau + \del_\sigma) - r_2(\del_\psi + \del_\phi) + r_3 \del_t - r_4 \del_y \,,
\end{aligned}
\end{align}
and so we obtain the one-forms $\theta_{a}$,
\begin{align}
\begin{aligned}
\label{eq: theta1forms}
	\theta_{\LLL} & = -n_5 \left[
	  \left( \cosh^2 \rho \, d\tau + \sinh^2\rho \, d\sigma  
	\right) +
	l_2 \left( \cos^2 \theta \, d\psi - \sin^2\theta \, d\phi  
	\right) \right] - ( l_3 \, dt + l_4 \, dy )
	\, , \\
	\theta_{\RRR} & = ~\,  n_5 \left[
	 \left( \cosh^2 \rho \, d\tau - \sinh^2\rho \, d\sigma  
	\right) +
	r_2 \left( \cos^2 \theta \, d\psi + \sin^2\theta \, d\phi
	\right) \right] + r_3 \, dt + r_4 \, dy \, .
\end{aligned}
\end{align}
The full null-gauged Wess-Zumino-Witten action is then
\begin{equation}
	S = S_{0}^\sl + S_{\cA}^\sl + S_{0}^{\su} + S_{\cA}^\su +  S_{0}^{t,y} +  S_{\cA}^{t,y} \,,
\end{equation}
with
\begin{align}
	S_{0}^{\sl} & =  \frac{n_5}{\pi}  \int  \Big[ \del\rho \:\! \bdel \rho + \sh^2\rhoo \, \del \sigma \bdel \sigma - \ch^2\rhoo \, \del \tau \bdel \tau -\sh^2 \rhoo \left(\del \sigma \bdel\tau - \del \tau \bdel \sigma \right) \Big] d^2z \,, \nn \\
	S_{\cA}^\sl &=  \frac{2 \:\! n_5}{\pi} \int \Big[\,  \bar{\Aa} \left( \sh^2 \rhoo \, \del \sigma  + \ch^2 \rhoo \, \del \tau   \right) +  \Aa \left(  \ch^2 \rhoo \, \bdel \tau  -\sh^2 \rho \, \bdel \sigma \right) -  \Aa \bar{\Aa} \, \ch(2\rho) \Big]\; d^2z \,, \nn\\
	S_{0}^{\su}& =  \frac{n_5}{\pi} \; \int \Big[  \del \theta \bar{\del}\theta + c^2_\theta \del\psi \bar{\del}\psi + s^2_\theta \del\phi \bar{\del}\phi   +  c_{\theta}^2 \; (\del \phi \bdel \psi - \bdel \phi \del \psi) \Big] \;  d^2z \,, \nn\\
	S_{\cA}^\su &=  \frac{2 \:\! n_5 }{\pi} \int \Big[ l_2 \bar{\Aa} \left( c^2_\theta \, \del \psi -s^2_\theta \, \del \phi  \right) + r_2 \Aa \left( c^2_\theta \, \bdel \psi + s^2_\theta \, \bdel \phi  \right) + l_2 r_2 \Aa \bar{\Aa} \, \cos(2\theta)\Big] \; d^2z \,, \nn\\
	 S_{0}^{t,y}  &=  \frac{1}{\pi} \int \Big[-\del t \bdel t + \del y \bdel y \Big]   d^2 z\,,\cr  
	 S_{\cA}^{t,y} &= \frac{2}{\pi} \int \Big[  l_3 \bar{\Aa} \del t + r_3 \Aa \bdel t + l_4 \bar{\Aa} \del y + r_4 \Aa \bdel y - (l_3 r_3 - l_4 r_4)\Aa\bar{\Aa}       \Big]   d^2 z\,,
\end{align}
where we have used the shorthands  $c_{\theta}=\cos\theta$ and  $s_{\theta}=\sin\theta$. 

We note that with the Killing vectors \eqref{eq: killing vectors}, the quantity $\Sigma$ defined in \eqref{eq:sigma-def} becomes 
\begin{equation}
\label{eq:Sigma}
   \Sigma \;=\; \frac12 \Big( n_5 \big[ \:\!\! \cosh (2\rho) - l_2 r_2 \cos(2\theta)  \big] + l_3 r_3 - l_4 r_4 \Big) \,.
\end{equation}
For convenience let us define the rescaled quantity
\begin{align}
   \Sigma_0 \;&=\; \frac{1}{n_5} \Sigma \;=\; 
   \sinh^2 \rhoo
   - l_2 r_2 \cos^2 \theta +\frac{1+l_2 r_2}{2}
   + \frac{l_3 r_3 - l_4 r_4}{2 n_5}  \,.
\end{align}

\subsection{Supergravity fields}

By integrating out the gauge fields and choosing the gauge $\sigma = \tau = 0$, we obtain the following line element and $B$-field:
\begin{align}
\begin{aligned} \label{eq:metric-B-gen}
	ds^2  \;=\;  & {} -  \frac{h_t}{\Sigma_0} dt^2 + \frac{h_y}{\Sigma_0} dy^2 + \frac{(l_3 r_4 + l_4 r_3)}{n_5 \Sigma_0} dt dy  \cr
	&{} + n_5 (d\theta^2 + d\rho^2) + n_5  \, \frac{ h_\phi}{\Sigma_0} \, \sin^2 \theta d\phi^2 + n_5  \,  \frac{h_\psi}{\Sigma_0} \, \cos^2 \theta d\psi^2  \cr
	&{} - \frac{1}{\Sigma_0} \left[(l_2 r_3 - l_3 r_2) dt + (l_2 r_4 - l_4 r_2) dy   \right] \sin^2\theta d\phi  \cr
	&{} + \frac{1}{\Sigma_0} \left[ (l_2 r_3 + l_3 r_2) dt + (l_2 r_4 + l_4 r_2) dy   \right] \cos^2\theta d\psi \,, \\[4mm]  
	B \;=~ & {} \frac{(l_3 r_4 - l_4 r_3)}{2 n_5 \Sigma_0} dt \wedge dy +  {n_5 \frac{h_\phi}{\Sigma_0} \cos^2 \theta} \;  d\phi \wedge d\psi  \cr
	& {} + \frac{ 1}{2 \Sigma_0} \left[ (l_2 r_3 + l_3 r_2) dt + (l_2 r_4 + l_4 r_2) dy   \right] \wedge \sin^2\theta d\phi  \cr
	& {} -  \frac{ 1}{2 \Sigma_0} \left[ (l_2 r_3 - l_3 r_2) dt + (l_2 r_4 - l_4 r_2) dy   \right] \wedge \cos^2\theta d\psi \,,
\end{aligned}
\end{align}
where
\begin{align}
\begin{aligned} \label{eq:hs}
	h_t & = \sinh^2 \rhoo
   - l_2 r_2 \cos^2 \theta 
   +\frac{1+l_2 r_2}{2}
   - \frac{l_3 r_3 + l_4 r_4}{2 n_5}
    \,, \\
	h_y & = \sinh^2 \rhoo
   - l_2 r_2 \cos^2 \theta +\frac{1+l_2 r_2}{2}
   + \frac{l_3 r_3 + l_4 r_4}{2 n_5} \,, \\
	h_\phi & = \sinh^2 \rhoo
	+\frac{1-l_2 r_2}{2}
   + \frac{l_3 r_3 - l_4 r_4}{2 n_5}
   \,,\\
	h_\psi & = \sinh^2 \rhoo +\frac{1+l_2 r_2}{2}
   + \frac{l_3 r_3 - l_4 r_4}{2 n_5}
   \,.
\end{aligned}
\end{align}

A non-trivial dilaton $\Phi$ is generated as usual at one-loop level on the worldsheet. 
In the null-gauging formalism, this arises from a change in the measure in the path integral formulation. The most direct way to compute the dilaton is by considering the usual one-loop beta function (equivalently the supergravity equations of motion).
This fixes $e^{2\Phi}$ to be proportional to
 \begin{equation}
 \label{eq:dilaton-pre}
      e^{2\Phi} \;\sim\; \frac{1}{\Sigma_0}\, .
 \end{equation}
 
The overall normalization of the dilaton can be fixed by matching to the NS5-brane decoupling limit of known solutions; we shall discuss this in detail in Section~\ref{sec:JMaRTuniqueness}. Nevertheless, let us make some preliminary comments on this in order to highlight the physical meaning of this constant. The simplest scenario corresponds to the solution sourced by a stack of $n_5$ coincident fivebranes, which is described by using the harmonic function 
 \begin{equation}
 \label{H5original}
H_5 \;=\; 1+\frac{n_5}{r^2} \,,
 \end{equation}
where $r$ is a radial coordinate, and where the dilaton is given by $e^{2\Phi}=g_s^2 H_5$. The fivebrane decoupling limit corresponds to $g_s \to 0$ with fixed $r/g_s$ and fixed $\alpha'$~\cite{Kutasov:2001uf} (recall that we have set $\alpha'=1$), which can be implemented via a scaling limit $g_s\to\epsilon$, $r \to \epsilon r$, with $\epsilon\to 0$. This brings the dilaton to the form 
 \begin{equation}
e^{2\Phi} \;=\; \frac{n_5}{r^2} \,.
 \end{equation}
Here we could have kept a fiducial rescaled $\tilde{g}_s$ (i.e.~$g_s=\epsilon \:\! \tilde{g}_s$) as in \cite{Martinec:2017ztd,Martinec:2018nco}, but since the asymptotic value of $e^{2\Phi}$ is zero this has no precise physical meaning. 
Next, for an array of $n_5$ fivebranes in a circular, $\mathbb{Z}_{n_5}$ symmetric configuration, the supergravity solution sees a smeared source and the relevant harmonic function is based on the function $\tilde\Sigma=r^2 +a^2 \cos^2\theta$, where the scale $a$ parametrises the radius of the circular array (see e.g.~\cite{Martinec:2017ztd} and references within). In this case we take a double scaling limit given by $g_s \to 0$ with fixed $r/g_s$, fixed $a/g_s$ and fixed $\alpha'$~\cite{Giveon:1999px,Giveon:1999tq}, which can be implemented via a scaling limit $g_s\to\epsilon$, $r \to \epsilon r$, $a \to \epsilon a$ with $\epsilon\to 0$. Changing variables to $r=a\sinh\rho$ in order to match the notation used above, we have $\tilde\Sigma=a^2(\sinh^2\rhoo+\cos^2\theta)\equiv a^2\tilde\Sigma_0$. The harmonic function in \eqref{H5original} is replaced by~\cite{Sfetsos:1998xd,Israel:2004ir,Martinec:2017ztd}
 \begin{equation}
H_5 \;=\; 1+\frac{n_5}{a^2\tilde{\Sigma}_0} \, ,
 \end{equation}
 so in the decoupling limit the dilaton takes the form 
 \begin{equation}
e^{2\Phi} \;=\; \frac{n_5}{a^2\tilde\Sigma_0} \,.
 \end{equation}

The backgrounds we consider will turn out to be generalisations of the circular array of fivebranes, such that, as a general expectation, the normalisation constant for the exponentiated dilaton in Eq.~\eqref{eq:dilaton-pre} should be proportional to the number of NS5 branes in the geometry. When F1 charge is also present this gets divided by $n_1$, giving a factor $n_5/n_1$ (or $\sim n_5/n_p$ in the NS5-P frame). Furthermore, there should also be a factor in the denominator given by the square of a length scale characterising the distribution of the sources. 
There will turn out be two lengthscales $a_1$, $a_2$ generalizing the scale $a$, and the decoupling limit involves scaling $g_s\to\epsilon$, $r \to \epsilon r$, $a_1 \to \epsilon a_1$, $a_2 \to \epsilon a_2$ with $\epsilon\to 0$~\cite{Martinec:2018nco}.
At this point however, we are working generally, so we do not yet know the details of the underlying bound state of branes. We postpone the precise computation until Section \ref{sec:JMaRTuniqueness}.

Together with the constraints \eqref{eq: null gauge constraints} on the $l_i, r_i$ parameters, the expressions for the supergravity fields \eq{eq:metric-B-gen}, \eq{eq:hs}, \eq{eq:dilaton-pre} describe the most general backgrounds that can be obtained within the class of null-gauged models considered in this paper, under our assumption $\lL_1\neq 0$, $\rR_1\neq0$. Models in which  $\lL_1=0$ or $\rR_1=0$ can easily be treated as a special case and we shall not consider them further.

As mentioned above, it is known that these models include the JMaRT solutions and their limits \cite{Martinec:2017ztd,Martinec:2018nco}. In this paper we shall prove that these are in fact all consistent solutions in this class of null-gauged models. Moreover, we will show that this conclusion can be reached either from consistency of the worldsheet CFT or from asking that the supergravity fields \eq{eq:metric-B-gen}, \eq{eq:hs}, \eq{eq:dilaton-pre} are free of CTCs, horizonless and smooth up to physical sources of string theory (in our cases, orbifold singularities or NS5-brane singularities).

\section{Consistency of the worldsheet
spectrum}
\label{sec:CFTanalysis}

In the previous section we introduced a class of null-gauged models from a classical point of view. Here we discuss the corresponding worldsheet coset CFTs, focusing on the relevant algebraic considerations and the associated spectrum. We start by briefly reviewing the construction of superstring theory on $AdS_3 \times \SSS^3 \times \TT^4$ generated by $n_5$ NS5 branes and $n_1$ fundamental strings. This is a pure NSNS background, which can be treated from the worldsheet perspective as a WZW model based on the group manifold $SL(2,\R)\times SU(2) \times U(1)^4$. Then, 
we introduce the novel ingredients of null-gauged models. We discuss how the BRST charges are modified and under which conditions the resulting background is supersymmetric. Finally, we derive a series of constraints leading to a consistent gauge-invariant spectrum. Spectral flow considerations play a key role in the analysis below.

\subsection{Brief review of superstrings in $AdS_3 \times \SSS^3 \times \TT^4$}
\label{sec:STADS3}

The $SL(2,\R)$ WZW model was studied in detail in \cite{Maldacena:2000hw,Maldacena:2000kv,Maldacena:2001km}. Here we will follow the notation of \cite{McElgin:2015eho}. The $SL(2,\R)$ currents  satisfy the OPEs
\begin{equation}
    j^a(z)j^b(w) \sim \frac{\eta^{ab} k/2}{(z-w)^2} + \frac{ f^{ab}_{\phantom{ab}c} j^c(w)}{z-w} 
    \label{OPEjSL2}
\end{equation}
where $k$ is the level of the affine algebra, while 
$-2\eta^{33} = \eta^{+-} = 2$,$f^{+-}_{\phantom{+-}3}=-2$ and $
    f^{3+}_{\phantom{3+}+}=-
    f^{3-}_{\phantom{3-}-}=1
$. The energy-momentum tensor and the central charge follow from the Sugawara construction and are given by
\begin{equation}
    T_{\mathrm{sl}}(z) = \frac{1}{k-2} :-j^3 (z) j^3(z) + \frac{1}{2} \left[j^+(z) j^-(z) + j^-(z) j^+(z)\right] :,  
    \label{TSL2}    
\end{equation}
and 
\begin{equation}
    c_{\mathrm{sl}} = \frac{3k}{k-2},
\end{equation}
respectively. Identical expressions hold for the anti-holomorphic sector.
    
The canonical spectrum of the model is built out of lowest- and highest-weight, and continuous representations of the zero-mode algebra. A principal discrete series of lowest weight is built out of the state $|j,j\rangle$, annihilated by $j_0^-$ by acting with $j_0^+$, thus spanning 
\begin{equation}
    \label{D+rep}
        {\cal{D}}_j^+ =
        \Big\langle \, 
        |j,m\rangle \ , \ m=j
        ,j+1,j+2,\cdots
        \Big\rangle,
    \end{equation}
where $j_0^3|j,m\rangle = m|j,m\rangle$. This is a unitary representation for any $j$ real and positive. The corresponding conjugates ${\cal{D}}_j^-$ are highest-weight representations, defined analogously.  
For consistency one must restrict to 
\begin{equation}
        \frac{1}{2} < j < \frac{k-1}{2},
        \label{Djrange}
\end{equation}
as follows from $L^2(AdS_3)$ normalisation conditions, no-ghost theorems and spectral flow considerations to be discussed below. On the other hand, principal continuous series are 
\begin{equation}
\label{Contrep}
    {\cal{C}}_j^\alpha = 
        \Big\langle \,
        |j, m, \alpha\rangle \ , \ 0 \leq \alpha < 1 \ , \ j=\frac{1}{2} + i s \ , \ s \in \R \ , \ m=\alpha
        ,\alpha\pm 1,\alpha \pm 2,\cdots
        \Big\rangle.
    \end{equation}
All states in \eqref{D+rep} and \eqref{Contrep} give rise to primary fields with conformal weights given by 
\begin{equation}
    \Delta = -\frac{j(j-1)}{k-2}. 
\end{equation}

A spectral flow automorphism of the current algebra is defined as
\begin{equation}
       j^\pm(z) \to \tilde{j}^\pm(z) = z^{\pm w} j^\pm(z) \ , \ 
    j^3(z) \to \tilde{j}^3(z) = j^3(z) -\frac{k\w}{2} z^{-1} ,
    \label{sl2bosflow2}
\end{equation}
where the so-called spectral flow charge $\w$ is an integer number. This induces, in turn, an automorphism of the Virasoro algebra given by  
\begin{equation}
\label{LntildeSL2}
    L_n \to \tilde{L}_n = L_n + \w j_n^3 - \frac{k}{4}\w^2 \delta_{n,0},
\end{equation}
where $L_n$ denotes the modes of \eqref{TSL2}. Analogous formulas hold for the anti-holomorphic sector. We work with the universal cover of $SL(2,\R)$, which further imposes that the left and right spectral flows must be equal, namely $\bar{\w}= \w$.

As it was shown in \cite{Maldacena:2000hw}, the action of \eqref{sl2bosflow2} on the canonical affine representations discussed above defines, in general, inequivalent representations that must be considered in order to generate a consistent spectrum. An exception occurs, however, given that the module obtained by flowing the affine representation ${\cal{D}}_j^+$ in $\w$ units is identical to that obtained by flowing  ${\cal{D}}_{k/2-j}^-$ in $\w-1$ units. The spectrum is thus constructed solely upon lowest-weight representations with $j$ restricted to the range \eqref{Djrange}.

Spectrally flowed primary states are not affine primaries. They are, however, Virasoro primaries with weight
\begin{equation}
    \Delta = -\frac{j(j-1)}{k-2} - m \w - \frac{k}{4}\w^2,
    \label{defDeltaw}
\end{equation}
as follows from \eqref{LntildeSL2}.

\medskip 

The supersymmetric affine $\widehat{\sl(2,\RR)}_{n_5} $ algebra is generated by the supercurrents $ \psisl^a + \theta \:\! J^a$, where $\theta$ is a formal Grassmann variable. The currents $J^a$ satisfy \eqref{OPEjSL2} with level $n_5$, and the OPEs involving the fermions are 
\begin{align}
	J^a(z) \psisl^b(w) & \sim  \frac{ f^{ab}_{\phantom{ab}c}\psisl^c(w)}{(z-w)}, 
	\label{JpsiOPE} \\
	\psisl^a(z) \psisl^b(w) & \sim \frac{\frac{n_5}{2} \, \eta^{ab}}{(z-w)}.
\end{align}
One can split the $J^a$ currents into two independent contributions as
\begin{align}
	J^a & = j^a - \frac{1}{n_5}f^{a}_{\phantom{a}bc} \psisl^b \psisl^c.
\end{align}
The bosonic currents $j^a$ generate an $ \widehat{\sl(2,\R)}_k $ algebra  with level $k=n_5+2$, commuting with the free fermion system. In the fermionic sector, the spectral flow automorphisms are given by \begin{equation}
    \psisl^\pm(z) \to \tilde{\psisl}^\pm(z)  = z^{\pm \w} \psisl^\pm(z) \ , \ 
	\psisl^3(z) \to \tilde{\psisl}^3(z)  = \psisl^3(z) 
	\label{flowfermionmodesSL2}
\end{equation}
while the corresponding maps for the $J^a$ and $j^a$ currents are as in \eqref{sl2bosflow2} with the respective levels. 

\medskip

The bosonic WZW model based on the $SU(2)$ group manifold was studied in \cite{Zamolodchikov:1986bd,Fateev:1985mm}. The generators of the current algebra will be denoted $k^a$, and for most quantities we use primes to distinguish them from their $SL(2,\R)$ counterparts. They satisfy the OPEs
\begin{equation}
    k^a(z)k^b(w) \sim \frac{\delta^{ab} k'/2}{(z-w)^2} + \frac{ f'^{ab}_{\phantom{ab}c} k^c(w)}{z-w}, 
    \label{OPEkSU2}
\end{equation}
where $k'$ is the level of the affine Lie algebra, $\delta^{ab}$ is the Killing form and $f'^{abc}$ are the corresponding structure constants, namely $2\delta^{33} = \delta^{+-} = 2$, $ f'^{+-}_{\phantom{+-}3}=2$, $
    f'^{3+}_{\phantom{3+}+}=-
    f'^{3-}_{\phantom{3-}-}=1$.
The energy momentum tensor is
\begin{equation}
    T_{\mathrm{su}}(z) = \frac{1}{k'+2} :\!\!\:k^3 (z) k^3(z) + \frac{1}{2} \left[k^+(z) k^-(z) + k^-(z) k^+(z)\right] \!\!\: : .
\end{equation}
This gives the central charge 
\begin{equation}
    c_{\su} = \frac{3k'}{k'+2}.
\end{equation}
The unitary representations of the zero-mode algebra upon which the $SU(2)_{k'}$ WZW spectrum is constructed are labelled by 
\begin{equation}
    j'  \in \Z/2 \ , \quad~~ 0\leq j' \leq k'/2,
\end{equation}
and they are spanned by 
\begin{equation}
    |j',m'\rangle \ , \ m'= -j',-j'+1,\dots, j'-1,j',
\end{equation}
where $m'$ is the eigenvalue of $k_0^3$. 
The associated primary fields have weights
\begin{equation}
    \Delta' = \frac{j'(j'+1)}{k+2}.
\end{equation}

Unlike in the $SL(2,\R)$ case, for the $SU(2)$ WZW model spectral flow is not necessary for constructing a consistent spectrum due to the compactness of the underlying manifold. Indeed, the spectral flow automorphisms merely reshuffle primary and descendant fields, and they do not introduce new inequivalent representations. Nevertheless, for superstring theory applications (and in the context of the null-gauged models in particular) it is of practical use to include it in the discussion\footnote{For instance, $SU(2)$ spectral flow is a convenient way to describe specific combinations of excitations such as giant gravitons.} \cite{Israel:2004ir,Giribet:2007wp,Martinec:2018nco,Martinec:2020gkv}. 
In the $SU(2)_{k'}$ context, spectral flow is defined as 
\begin{equation}
    k^\pm(z) \to \tilde{k}^\pm(z) = z^{\mp w'} k^\pm(z) \ , \ 
    k(z)^3 \to \tilde{k}^3(z) = k^3(z) - \frac{k'\w'}{2} z^{-1}.
    \label{su2bosflow2}
\end{equation}
The associated shift of the Virasoro modes is 
\begin{equation}
    L_n \to \tilde{L}_n = L_n - \w' k_n^3 + \frac{k'}{4}\w'^2 \delta_{n,0}.
\end{equation}
Similar expressions hold for the anti-holomorphic sector. In this case, however, 
it is possible to have $\bar{\w}'\neq \w'$.
As before, spectrally flowed primary fields are Virasoro primaries of weight
\begin{equation}
    \Delta' = \frac{j'(j'+1)}{k'-2} + m' \w' + \frac{k'}{4}\w'^2 \,,
    \label{defDeltapw}
\end{equation}
however they are not affine primaries. 

\medskip

As for the $SL(2,\R)$ case, the supersymmetric $SU(2)_{n_5}$ generators are given by  
$ \psisu^a + \theta K^a$, where the currents $K^a$ satisfy \eqref{OPEkSU2} with the same level $n_5$, while the rest of the OPEs take the form  
\begin{align}
	K^a(z) \psisu^b(w) & \sim  \frac{f'^{ab}_{\phantom{ab}c}\psisu^c(w)}{(z-w)} , 
	\label{KopeCHI}\\
	\psisu^a(z) \psisu^b(w) & \sim \frac{\frac{n_5}{2} \, \delta^{ab}}{(z-w)} .
\end{align}
As before, it is convenient to define
\begin{align}
	K^a & =\sucur^a - \frac{1}{n_5}f'^{a}_{\phantom{a}bc} \psisu^b \psisu^c,
\end{align}
where now the bosonic currents $k^a$ generate an $ \widehat{\su(2)}_{k'} $ algebra with level $k'=n_5-2$. The spectrally flowed fermionic modes are analogous to the $SL(2,\R)$ case, i.e.
\begin{equation}
    \psisu^{\pm}(z) \to \tilde{\psisu}^\pm(z)  = z^{\mp w}\psisu^\pm(z) 
    \ , \quad~~ 
	\psisu^3 (z) \to \tilde{\psisu}^3(z)  = \psisu^3 (z) . 
	\label{flowfermionmodesSU2}
\end{equation}

We can now describe the worldsheet theory for superstrings on $AdS_3\times \SSS^3 \times \TT^4$. 
The energy momentum tensor $T$ and supercurrent $G$ of the full WZW model are given by
\begin{eqnarray}
    T &=& \frac{1}{n_5} \left(j^a j_a - \psisl^a \der \psisl_a + 
    k^a k_a - \psisu^a \der \psisu_a 
    \right) + \frac{1}{2}
    \left(\der Y^i \der Y_i - \lambda^i \der \lambda_j\right),
    \label{TAdS3S3T4def}
    \\
    G &=& \frac{2}{n_5} \left(
    \psisl^a j_a - \frac{1}{3n_5} f_{abc} \psisl^a \psisl^b \psisl^c + 
    \psisu^a k_a - \frac{1}{3n_5} f'_{abc} \psisu^a \psisu^b \psisu^c
    \right) + \lambda^i \der Y_i,
    \label{GAdS3S3T4def}
\end{eqnarray}
where the flat compact directions are treated as usual by including  four canonically normalized free bosons $Y^i$ and their fermionic partners $\lambda^i$ $(i=6,\dots,9)$. 
The (matter) central charge is 
\begin{equation}
    c = \frac{3 (n_5+2)}{n_5} + \frac{3}{2} + \frac{3 (n_5-2)}{n_5} + \frac{3}{2} + 6 = 15.
\end{equation}
In the superstring model this is cancelled out by the contribution coming from the usual $bc$ and $\beta \gamma$ ghost systems. The corresponding BRST charge is given by 
\begin{equation}
    {\cal{Q}} = \oint dz :c \left(T + T_{\beta\gamma}\right) - \gamma \, G + c(\der c) b - \frac{1}{4} b \gamma^2:,
\end{equation}
where $T_{\beta\gamma}$ is the energy-momentum tensor of the $\beta\gamma$ system.

It is useful to bosonize the latter as
\begin{align}
	\beta = e^{-\varphi} \del \xi\qqquad \gamma = \eta \:\! e^{\varphi},
\end{align}
where $\varphi$ is a canonically normalized scalar with background charge $2$, and $\xi(z)\eta(w) \sim (z-w)^{-1}$. The rest of the fermions are dealt with similarly by introducing  (c.f.~\cite{Giveon:1998ns,Itzhaki:2005tu,Martinec:2020gkv}) the canonically normalized bosonic fields $H_I$ with $I=1,\dots 5$ as 
\begin{gather}
	 i\del H_1 = -\frac{2i}{n_5} \, \psisl^1\psisl^2 = \frac{1}{n_5} \, \psisl^+\psisl^- \, , \qquad  i \del H_2 = -\frac{2i}{n_5} \, \psisu^1\psisu^2 =  \frac{1}{n_5} \, \psisu^+\psisu^-  \, ,
	 \nonumber \\ 
	 i \del H_3 = \frac{2}{n_5} \, \psisl^3\psisu^3  \, ,\qquad 
	 i\del H_4 = i\lambda^6 \lambda^7 \, , \qquad i\del H_5 = i\lambda^8 \lambda^9 \, ,
\end{gather}
where $ H_I^\dagger = H_I$ for $I \ne 3 $ and $ H_3^\dagger = - H_3 $. In terms of these, the space-time supercharges read 
\cite{Giveon:1998ns}: 
\begin{equation}
    Q_\vep = \oint dz \, e^{-\varphi/2} S_\vep \ , \quad \ S_\vep = \exp \left(\frac{i}{2} 
    \sum_{I=1}^{5}\vep_I H_I\right),
    \label{supercharges}
\end{equation}
where $S_\vep$ are the spin fields and $\vep_I=\pm 1$.
One can show that BRST invariance and mutual locality of the supercharges respectively impose the conditions
\begin{equation}
    \prod_{I=1}^{3} \vep_I = \prod_{I=1}^{5} \vep_I = 1.
\end{equation} 
The eight possible sign choices then define the supercharges of the $\Nn=4$ superconformal spacetime algebra.

Notice that, according to Eqs.~\eqref{flowfermionmodesSL2} and \eqref{flowfermionmodesSU2}, the respective spectral flow operations in $SL(2,\R)$ and $SU(2)$ act on fermion fields as  
\begin{equation}
    \tilde{\psisl}^\pm = \psisl^\pm e^{- i \w H_1} \ , \qquad\quad 
    \tilde{\psisu}^\pm = \psisu^\pm e^{ i \w' H_2},
    \label{flowH12}
\end{equation}
while the rest of the fermions remain unchanged.

\medskip 

We now discuss the physical states, focusing on the NSNS sector. We briefly review the differences between discrete and continuous representations, and also the consequences of including spectral flow. 
The unflowed sector was discussed in \cite{Kutasov:1998zh}.
Here, the ground state is tachyonic as it does not survive the GSO projection. On the other hand, for states with one fermionic excitation one combines bosonic primaries and free fermions into spin $j\pm 1$ and $j'\pm 1$ representations of the supersymmetric currents. 
The Virasoro condition for these states reads 
\begin{equation}
 \frac{1}{2} + \frac{1}{2} - \frac{j(j-1)}{n_5} + 
 \frac{j'(j'+1)}{n_5} = 1, 
 \label{Virasoro}
\end{equation}
which is solved by setting $j'=j-1$, so that one is restricted to deal with the discrete representations. 
By identifying the space-time theory weight and charge, one recognises chiral multiplets of the holographic CFT. 

Spectrally flowed states were constructed for instance in Refs.~\cite{Argurio:2000tb, Hofman:2004ny} for generic $AdS_3\times {\cal{N}}$ models, and also in \cite{Giribet:2007wp,Eberhardt:2019qcl} in the present context. In the supersymmetric theory, the $SL(2,\R)$ contribution to the conformal weight of a spectrally flowed  primary takes the form 
\begin{equation}
    \Delta = -\frac{j(j-1)}{n_5} - m \w - \frac{n_5}{4} \w^2.
\end{equation} 
This is indeed the case once the extra factor $e^{i \w H_1}$ associated with the shift of the fermionic modes in \eqref{flowfermionmodesSL2} is included (see Eq.~\eqref{flowH12}). More precisely, the spectral flow operator is given by 
\begin{equation}
    \exp \left(-i \w H_1 + \w \sqrt{\frac{n_5+2}{2}} \phi
    \right),
    \label{susyflowoperator}
\end{equation}
where $\phi$ bosonizes the current $j^3$. By flowing primaries with no excitations in this way, we obtain states that are actually consistent with GSO for odd values of $\w$, and the corresponding Virasoro condition becomes
\begin{equation}
    \frac{1}{2} - \frac{j(j-1)}{n_5} - m\w - \frac{n_5}{4}\w^2 + \frac{j'(j'+1)}{n_5}=1,
    \label{virasorocontinuous}
\end{equation}
while for singly excited states we need $\w$ to be even, and also to add an extra $1/2$ on the LHS of the previous equation. 
When the unflowed bosonic $SL(2,\R)$ primary belongs to the continuous representation, we have $j = \frac{1}{2} +i s$ with $s  \in \R$ such that $-j(j-1) = \frac{1}{4}+s^2$, hence we use \eqref{virasorocontinuous} to solve for $m$ (with $s$ and $j'$ fixed). The vertex operators obtained in this way constitute the long string sector, which was recently interpreted in \cite{Eberhardt:2019qcl} in terms of a Liouville factor included in the seed theory of the symmetric orbifold holographic CFT.

In the discrete case, $m$ can only take values $\pm (j + n)$ with $n \in \mathbb{N}_0$, and, as a consequence, the modified Virasoro condition becomes difficult to solve in general.
A simple way out is to further include spectral flow in the $SU(2)$ sector, generating additional terms as in \eqref{defDeltapw} (again with $n_5$) which can be used to cancel the extra $SL(2,\R)$ contributions if $\w' = \w$. More precisely, we use a spectral flow operator of the form
\begin{equation}
    \exp \left(-i \w H_1 + \w \sqrt{\frac{n_5+2}{2}} \phi
    + i \w H_2 + i \w \sqrt{\frac{n_5-2}{2}} \phi'\right), \label{susyflowoperator2}
\end{equation}
which, moreover, is mutually local with the supercharges \eqref{supercharges}, ensuring that flowed excited states will also survive the GSO projection. Instead of \eqref{virasorocontinuous}, the Virasoro condition for such spectrally flowed states becomes 
\begin{equation}
 \frac{1}{2} + \frac{1}{2} - \frac{j(j-1)}{n_5} - m \w - \frac{n_5}{4}\w^2 
 + \frac{j'(j'+1)}{n_5} + m'\w +\frac{n_5}{4}\w^2  \;=\; 1.
 \label{virasoroflow2}
\end{equation}
Once again, we impose $j'=j-1$ and are thus left with the condition $m=m'$. This is a highly restrictive condition that only specific highest/lowest-weight operators satisfy. We also note that the associated values for $m$ and $m'$ ensure that these flowed states are BRST invariant \cite{Giribet:2007wp}. Furthermore, they correspond to additional chiral states of the holographic CFT.

Of course, at the beginning of the discussion leading to \eqref{susyflowoperator2} we could have chosen to work with opposite spectral flows on the $SU(2)$ sector, i.e. $\w' = -\w$. This alternative spectral flow operator would lead to $m'=-m$ instead. As was discussed in \cite{Argurio:2000tb, Hofman:2004ny}, this is purely conventional: the set of operators obtained in this way is equivalent to the one we have described above.  

We note in passing that the requirement of introducing spectral flow in $SU(2)$ has been recently highlighted in the context 
of the tensionless string in $AdS_3$ \cite{Eberhardt:2018ouy}. This corresponds to the $n_5=1$ case, where the usual RNS formalism fails since the level of the bosonic $SU(2)$ factor  $k'=n_5-2$ would be negative. The hybrid formalism used in \cite{Eberhardt:2018ouy} makes use of the WZW model for the supergroup $PSU(1,1|2)$, whose maximal bosonic subgroup is $SL(2,\R)\times SU(2)$. In this context, the spectral flow operation is naturally carried out simultaneously on both factors with identical charges as in \eqref{susyflowoperator2}.

Finally, let us recall that all the above considerations concerning the spectrum were circumscribed to the holomorphic sector. Similar arguments hold for the antiholomorphic sector as well, thus imposing $\bar{\w}'=\w'=\w$. Consequently, the $AdS_3\times \SSS^3 \times \TT^4$ spectrum only contains states which have the same left and right $SU(2)$ spectral flow charge. However, we stress that this constraint will be relaxed for the null-gauged models analysed in the following sections. Indeed, the need for including operators with $\bar{\w}' \neq \w'$ was instrumental in the particular case of NS5 branes on the Coulomb branch studied in \cite{Israel:2004ir} (see also \cite{Martinec:2020gkv}).

\subsection{Superstring Theory in null-gauged models}

We now proceed to analyse the class of gauged WZW models introduced in Section \ref{sec:review}. In this section we work directly at the level of the coset CFT. We will see that a number of consistency conditions can be derived, which restrict the possible values of the parameters $l_i,r_i$ that define the embedding of the abelian subgroups being gauged. 

As advertised above, the transformations we gauge are chiral and correspond to $g \to h_L \, g \, h_R^{-1}$, where $g \in \Gg = SL(2,\R) \times SU(2) \times \R_t \times U(1)_y$ and $h_{L(R)} \in \Hh_{L(R)} = U(1)_{L(R)}$. Keeping the notation general, we have seen that introducing two independent gauge fields $\Aa,\bar{\Aa}$ transforming as 
\begin{equation}
    \Aa \to h_L\Aa \, h_L^{-1} + \der h_L h_L^{-1} \ , \ 
    \bar{\Aa} \to h_R \bar{\Aa} \, h_R^{-1} + 
    \bar{\der} h_R h_R^{-1} 
\end{equation}
leads to the gauge-invariant action \cite{Chung:1992mj}
\begin{equation}
    S[g,\Aa,\bar{\Aa}] = S_{\mathrm{WZW}}(g) + \frac{k}{\pi} \int d^2z \, \Tr \left[
    \Aa g^{-1}\bar{\der}g - \bar{\Aa} \der g g^{-1} - g^{-1}\bar{\Aa} g \Aa
    \right]. 
\end{equation}
By parametrising the gauge fields as 
\begin{equation}
\label{changeAhLR}
    \Aa = \der H_L H_L^{-1} \ , \
    \bar{\Aa} = \bar{\der} H_R H_R^{-1} \ , \ 
    \left(
    H_L \to h_L H_L \ , \ H_R \to h_R H_R
    \right) 
\end{equation}
and making use of the Polyakov-Wiegmann identity
\begin{equation}
    S_{\mathrm{WZW}}(a b) = S_{\mathrm{WZW}}(a) + S_{\mathrm{WZW}}(b) + \text{sgn}(\kappa) \frac{k}{\pi} 
    \int d^2z \, \Tr \left[
    a^{-1}\bar{\der}a \der b b^{-1}
    \right],
\end{equation}
we can rewrite the gauged action as 
\begin{equation}
    S[g,\Aa,\bar{\Aa}] = S_{\mathrm{WZW}}\left(H_L^{-1}gH_R\right) - 
    S_{\mathrm{WZW}}\left(H_L^{-1}\right) - 
    S_{\mathrm{WZW}}\left(H_R\right).
\end{equation}

A crucial simplification occurs when the currents being gauged are null. In this case we have $S_{\mathrm{WZW}}\left(H_L^{-1}\right) = S_{\mathrm{WZW}}\left(H_R\right) = 0$. Moreover, the Jacobian associated to the change of variables in Eq.~\eqref{changeAhLR} can be seen to trivialise for the same reason, except for the appearance of the usual $\tilde{b}\tilde{c}$ system \cite{Chung:1992mj}.
Finally, one can change variables from $g$ to the gauge-invariant $\Gg$-valued quantity $\tilde{g} = H_L^{-1}gH_R$. As a result, the path integral of the gauged theory is simply interpreted as that of the original upstairs WZW model on $\Gg$ combined with the ghost contributions. The same holds for the supersymmetric case. 

The consequences of the presence of the ghosts signalling the null gaugings can be understood intuitively as follows. Upon quantisation, we find that, on top of the usual string theory contributions to the BRST charges ${\cal{Q}}$ and $\bar{{\cal{Q}}}$, it becomes necessary to include new chiral terms of the (schematic) form 
\begin{equation}
\oint dz :\tilde{c} \, J:  \; , \quad  \  \oint dz :\bar{\tilde{c}} \, \bar{J}: \; ,    
\end{equation}
together with their fermionic counterparts.
This ensures that, under the gauging procedure outlined above, the spectrum of the coset model is built simply out of the vertex operators of the upstairs theory that are BRST-closed. In other words, physical operators must be gauge invariant.

\medskip 

We can now make this construction explicit for the models under consideration.  
These include black hole microstate solutions with up to three charges, and we view them in the NS5-F1-P duality frame, where the relevant fields are the metric, the $B$-field and the dilaton. The corresponding geometries were described in Section \ref{sec:review}. We know from~\cite{Martinec:2017ztd} that these models include both BPS and non-BPS black hole microstates.
Given that all the necessary ingredients belong to the NSNS sector we expect to have a well-defined solvable worldsheet model describing strings propagating in these backgrounds. Indeed, as shown in \cite{Martinec:2017ztd,Martinec:2018nco} and reviewed in Section \ref{sec:review} above, the worldsheet theory associated to the propagation of strings in this context corresponds to a coset CFT of the form
\begin{equation}
\frac{SL(2,\R) \times SU(2)  \times \R_t \times U(1)_y}{U(1)_L \times U(1)_R} \; \times U(1)^4 \,.
\label{cosetJMaRT}
\end{equation}

Let us first characterise the upstairs twelve-dimensional model. Here we simply add the extra time direction $t$ and spatial circle $y$ to the matter content employed in the previous section, together with the corresponding fermionic partners $\lambda^t$ and $\lambda^y$. The latter are bosonized as $i \der H_{6} = \lambda^t\lambda^y$, with $H_6^{\dagger} = - H_6$. They give additional free field contributions to the matter $T$ and $G$ in \eqref{TAdS3S3T4def} and \eqref{GAdS3S3T4def}.
The null current operators being gauged are then
\begin{align} 
\begin{aligned}
   J \,& =\, i{\cal{J}} \,=\, J^3 + l_2 K^3 + l_3 P_t + l_4 P_{y,L} \ , \ \quad \\
   \bar{J} \,& =\, i\bar{{\cal{J}}} \,=\, \bar{J}^3 + r_2 \bar{K}^3 + r_3 P_t + r_4 P_{y,R} \, , 
    \label{Jcurrents}
\end{aligned}
\end{align}
where $P_t = i \der t\,$, $P_{y,L} = i \der y\,$, and $P_{y,R} = i \bar\der y$. 
Together with the extra coordinates we also include the additional set of ghosts mentioned above, together with their fermionic partners. Note that it is necessary to take $h[\tilde{c}] = 0$ and $h[\tilde{\gamma}] = 1/2$, such that the central charges $c_{\tilde{b}\tilde{c}} = -2$ and $c_{\tilde{\beta}\tilde{\gamma}} = -1$ cancel the additional matter contribution $c_{ty} = 3$. This also implies that the bosonization
\begin{equation}
	\tilde{\beta} = e^{-\tilde{\varphi}} \del \tilde{\xi} \ , \quad~~  \tilde{\gamma} = \tilde{\eta} \, e^{\tilde{\varphi}}
\end{equation}
yields a canonically normalized scalar $\tvphi$ with no background charge. Consequently, we can work with $\tvphi$-independent vertex operators in the NSNS sector. On the other hand, the definition of the spin fields and the would-be spacetime supercharges is modified to \cite{Martinec:2020gkv}
\begin{equation}
    Q_\vep = \oint dz \, e^{-\left(\varphi - \tvphi\right)/2} S_\vep \ , \qquad S_\vep = \exp \left(\frac{i}{2} 
    \sum_{I=1}^{6}\vep_I H_I\right),
    \label{supercharges2}
\end{equation}
where the contributions to the conformal dimension of the integrand in $Q_\vep$ from the $\tvphi$ and $H_6$ exponentials cancel exactly. Note that the mutual locality condition now reads $\prod_{I=1}^6 \vep_I = 1$. An analogous formula defines the anti-holomorphic counterpart $\bar{Q}_\vep$.

As stressed above, the present procedure can lead to both BPS and non-BPS backgrounds. This depends on whether the charges $Q_\vep$ turn out to be BRST invariant or not, according to the precise current we choose to gauge. The left-handed BRST charge takes the form 
\begin{equation}
    {\cal{Q}} = \oint dz : \left[c \left(T + T_{\beta\gamma\tilde{\beta}\tilde{\gamma}}
    \right) + \gamma G + \tilde{c} {J} + \tilde{\gamma} \boldsymbol{\lambda} + \text{ ghosts } \right]: \, , 
    \label{BRSTcoset}
\end{equation}
and similarly for the right-handed one. Here $\boldsymbol{\lambda}$ and $\bar{\boldsymbol{\lambda}}$ are the superpartners of the currents in Eq.~\eqref{Jcurrents}, that is 
\begin{equation}
\label{fermioniccurrents}
    \boldsymbol{\lambda} = \psi^3 + l_2 \chi^3 + l_3 \lambda^t + l_4 \lambda^y 
    \ , \quad \ 
    \bar{\boldsymbol{\lambda}} = \bar{\psi}^3 + r_2 \bar{\chi}^3 + r_3 \bar{\lambda}^t + r_4 \bar{\lambda}^y.
\end{equation}
This ensures that only operators satisfying the usual Virasoro and $\gamma G$-invariance conditions that are also uncharged under the bosonic currents ${J}, \bar{J}$ and annihilated by $\tilde{\gamma} \boldsymbol{\lambda}$ and $\bar{\tilde{\gamma}} \bar{\boldsymbol{\lambda}}$ are physical. In particular, by using the $\gamma G$-invariance condition $\vep_1\vep_2\vep_3=-1$,  the $Q_\vep$ supercharges survive the gauging if and only if~\cite{Martinec:2020gkv}
\begin{equation}
    \vep_1 + \vep_2 l_2  = 0    \ , \quad \ 
    l_4 + \vep_6 \, l_3 = 0,
    \label{susyrestricparam}
\end{equation}
where the former constraint comes from the bosonic current and the latter arises from the fermionic one. Actually, only one of these two restrictions is independent due to the null condition \eqref{eq: null gauge constraints}. Analogously, for the antiholomorphic supercharges one has  
\begin{equation}
    \vep_1 + \vep_2 r_2  = 0    \ , \quad \ 
    r_4 + \vep_6 \, r_3 = 0,
    \label{susyrestricparamR}
\end{equation}
For instance, the cases with $(4,4)$ and $(4,0)$ spacetime supersymmetry were considered recently in \cite[App.~B]{Martinec:2020gkv}. In the present work we focus mainly on the more general non-supersymmetric case, in which there is no solution to this set of constraints since $l_2\neq \pm 1$ and $r_2 \neq \pm 1$.

We pause here to stress that we start from a (10+2)-dimensional model where all operators are taken to be  mutually local. This includes the charges $Q_\vep,\bar{Q}_\vep$. In particular,  this amounts to imposing the analogue of the GSO projection in the upstairs theory. Consequently, there are no tachyons in the resulting spectrum, nor in that of the coset model \cite{Martinec:2018nco,Martinec:2020gkv}. Indeed, even if no supersymmetry is preserved, the supergravity solutions we are dealing with are expected to be classically stable. Note that this is consistent with the fact that we only recover the fivebrane decoupling limit of JMaRT geometries, where there is no ergoregion~\cite{Martinec:2018nco}. The full solutions with flat asymptotics do exhibit the well-known ergoregion instability \cite{Cardoso:2005gj,Chakrabarty:2015foa}, interpreted as an enhanced version of Hawking radiation~\cite{Chowdhury:2007jx,Avery:2009tu,Avery:2009xr}.

\subsection{Physical spectrum and consistency conditions} \label{sec:consistency-CFT}

We now consider the physical states in the null-gauged theory. These are given in terms of vertex operators of the upstairs model that are BRST invariant as defined by the charge \eqref{BRSTcoset} and its anti-holomorphic counterpart. 

The lightest physical states (with no winding) are given by unflowed operators with a single fermionic excitation. All such operators must satisfy the Virasoro condition 
\begin{equation}
\label{nullgaugedVirasoro1}
   0 = - \frac{j(j-1)}{n_5} + \frac{j'(j'+1)}{n_5} 
    - \frac{1}{4} E^2 + \frac{1}{4} P_{y}^2 \,.
\end{equation}
These are automatically invariant under the action of $\boldsymbol{\lambda}$ in \eqref{fermioniccurrents}, so that they are BRST-invariant iff the following null-gauging\footnote{The factors of $2$ in the free field terms of Eqs.~\eqref{nullgaugedVirasoro1} and \eqref{nullgaugeconstr} arise from the OPEs $\partial t(z)\partial t(0)\sim -\frac12\frac{1}{z^2}$, $\partial y(z)\partial y(0)\sim \frac12\frac{1}{z^2}$ (recall that we work with $\alpha'=1$).} constraints hold:
\begin{equation}
\label{nullgaugeconstr}
0 = m + l_2 \, m' + \frac{l_3}{2} E + \frac{l_4}{2} P_{y} \ , \qquad
0= \bar{m} + r_2\, \bar{m}' + \frac{r_3}{2} E +\frac{r_4}{2} P_{y} \;.
\end{equation}
The simplest states of this sort are the 6D scalars,
whose spectrum was shown to coincide with that of minimally coupled massless scalar perturbations on top of the JMaRT geometries in \cite{Martinec:2018nco}. There are also the extremal-weight states, which for $E=P_y=0$ are half-BPS and were studied in \cite{Martinec:2020gkv} in the context of supertubes. Note that operators with $\TT^4$ polarisations can also give BPS states in specific models with non-trivial twisted sectors \cite{Kutasov:1998zh}.

On the other hand, operators with more general projections need to be combined with 
polarisations on the extra directions $t$ and $y$ in order to achieve BRST-invariance. The 
corresponding coefficients are determined by invariance under $\boldsymbol{\lambda}$ together with transversality in the $t,y$ directions.
The anti-holomorphic variables must satisfy analogous conditions as well, together with the gauge invariance conditions \eqref{nullgaugeconstr}.
Consequently, these constraints restrict the possible polarisations to those expected in the (9+1)-dimensional setting. 

Let us now consider spectrally flowed states. A particularly simple case corresponds to the circular array of 
fivebranes on the Coulomb branch, which is obtained by choosing $l_1 = l_2 = r_1 = - r_2 = 1$ and $l_3=l_4=r_3=r_4= 0$ \cite{Israel:2004ir, Martinec:2017ztd}. From the worldsheet point of view, this null-gauged model is analogous to the cigar construction used in \cite{Giveon:1999px, Giveon:1999tq}  provided we replace the extra circle by an $\SSS^3$ and take $J_y \to K^3$. Thus, as in that example, introducing spectral flow with $\w = \w' = \bar{\w}'$ does not produce new physical states since it merely amounts to a large gauge transformation. In a more general context, however, this is not true anymore, and spectrally flowed sectors contribute non-trivially to the spectrum.  
At this point, we also allow for both momentum $n_y$ and winding $\w_y$ on $\SSS^1_y$, and use the shorthands 
\begin{equation}
P_{y,L/R} =  \left(\frac{n_y}{R_y}\pm \w_y R_y \right)
\ , \ n_y,\w_y \in \mathbb{Z}.
\end{equation}
For generic states with spectral flow charges $\w$ on $SL(2,\R)$,  $(\w',\bar{\w}')$ on $SU(2)$, and winding $\w_y$ on $\SSS_y^1$, the null-gauge constraints \eqref{nullgaugeconstr} read
\begin{subequations}
\label{nullgaugeconstrW}
\begin{eqnarray}
    0 &=& 
    m + \frac{n_5}{2} \w +
    l_2 \left(
    m' + \frac{n_5}{2} \w'\right)
    + \frac{l_3}{2} E + \frac{l_4}{2} P_{y,L}\, , \\
    0 &=& 
    \bar{m} + \frac{n_5}{2} \w +
    r_2 \left(
    \bar{m}' + \frac{n_5}{2} \bar{\w}'\right)
    + \frac{r_3}{2} E + \frac{r_4}{2} P_{y,R},  
\end{eqnarray}
\end{subequations}
while the Virasoro constraints take the form
\begin{subequations}
\label{VirasoroNullWinding}
\begin{align}
    \frac{1}{2} &= 
    \frac{j'(j'+1)-j(j-1)}{n_5} - m \w
    + m' \w' + \frac{n_5}{4}\left(\w^{'2}-\w^2\right) - \frac{1}{4} \left(E^2 - P_{y,L}^2\right) + N
    \, , \\
    \frac{1}{2} &= 
    \frac{j'(j'+1)-j(j-1)}{n_5} - \bar{m} \w
    + \bar{m}' \bar{\w}' + \frac{n_5}{4}\left(\bar{\w}^{'2}-\w^2\right) - \frac{1}{4} \left(E^2 - P_{y,R}^2\right) +  \bar{N}
    .  
\end{align}
\end{subequations}
Here $N$ and $\bar{N}$ are the excitation numbers, and we have restricted to unflowed states with no fermion excitations for simplicity. 

The discussion so far does not characterise the physical spectrum in a unique way:
there is a residual discrete gauge orbit
connecting equivalent representatives of the same physical state. 
This fact was noticed in~\cite{Martinec:2018nco}, so let us first recall the observations made in that work before making a set of generalisations. First, spectral flow in the null direction corresponding to the gauge current is gauge-trivial. Second, the non-compactness of $\mathbb{R}_t$ means that 
there cannot be independent left and right gauge spectral flow transformations, since these would shift the zero mode of $t$ differently. Therefore, globally we work with the universal cover of $SL(2,\R)$, in which the left and right spectral flow parameters are constrained to be equal, $\omega=\bar{\omega}$. Moreover, globally we gauge $\mathbb{R}\times U(1)$, a (1+1)-dimensional cylinder composed of one compact spacelike direction and one non-compact timelike direction.
The gauged model then has a single non-compact timelike direction.\footnote{By contrast, in related models that do not include the $\R_t$ factor in the upstairs model, the single cover of $SL(2,\R)$ has been considered~\cite{Israel:2004ir}.}
Third, the non-compactness of the time coordinate $t$ moreover imposes
\begin{equation}
    \label{eq:l3equalr3}
    l_3 \,=\, r_3 \,,
\end{equation}
or, in terms of the original gauging parameters, $\lL_3/\lL_1=\rR_3/\rR_1$.
We will re-derive the condition $l_3=r_3$ from an independent point of view in the following section by imposing smoothness, absence of horizons, and absence of CTCs in the corresponding geometry.

We now make a more general analysis of this phenomenon in the general models defined in the previous section. 
 Let us stress that the analysis of such gauge orbits is not simply about the counting of states. Indeed, being able to identify gauge-equivalent operators in terms of the quantum numbers of the WZW model is necessary for building a consistent theory, and we will show that it further constrains the allowed values for the gauging parameters $l_i,r_i$.  

Given a physical state, let us seek spectral flow transformations that result in the same operator. 
By subtracting the two equations in \eq{VirasoroNullWinding} we find
\begin{equation}
\label{LevelMatching}
 0 =    \w (\bar{m} - m) + m' \w' - \bar{m}' \bar{\w}' + 
 \frac{n_5}{4}(\w^{'2}-\bar{\w}^{'2}) +
  n_y \w_y + N - \bar{N}, 
\end{equation}
which plays the role of the level-matching condition in this context. In order to find solutions of Eq.\;\eqref{LevelMatching},   $(\w^{'2}-\bar{\w}^{'2})$ must be a multiple of 4, so $\w' \pm \bar{\w}'$ must be even. Note that this preserves the statistics of the $SU(2)$ part of the state.

Let us consider a shift of the form\footnote{As in Section~\ref{sec:STADS3}, here we include both the bosonic and the free fermion spectral flows, so that the shifted charges and weights
are written in terms of the supersymmetric level $n_5$ as opposed to the bosonic levels $k=n_5+2$ and $k'=n_5-2$.} 
\begin{equation}
\label{shiftSL2}
 \w \to  \w + q \,, \qquad q \; \in \; \Z \,.
\end{equation}
We shall show that this can be compensated at the level of the null-gauge constraints \eqref{nullgaugeconstrW} without altering the weights \eqref{VirasoroNullWinding} by shifting the remaining quantum numbers appropriately. We begin with a general shift and show that only the shift in the null gauge direction achieves this.
We allow for arbitrary multiples of $q$ to shift $\w'$, $\bar{\w}'$, $E$, $n_y$ and $\w_y$ as well, namely 
\begin{equation}
    (\w',\bar{\w}',E,P_{y,L},P_{y,R}) \to 
    (\w'-a_2 q,\bar{\w}'-b_2q,E+a_3q,P_{y,L}-a_4q,P_{y,R}-b_4q).
\end{equation}
For the weights  \eqref{VirasoroNullWinding} and gauge constraint \eqref{nullgaugeconstrW} to remain unchanged for arbitrary $q$, we must have 
\begin{eqnarray}
    0 &=& m + \frac{n_5}{2} \w +
    a_2 \left(
    m' + \frac{n_5}{2} \w'\right)
    + \frac{a_3}{2} E + \frac{a_4}{2} P_{y,L}, \nn \\ [1ex]
    0 &=& n_5 (1-a_2^2)+a_3^2-a_4^2 , \\ [1ex]
    0 &=& n_5(1-l_2 a_2)+l_3 a_3-l_4a_4, \nn
\end{eqnarray}
and the same with $a_{2,4} \to b_{2,4}$, $l_{2,4}\to r_{2,4}$, $m'\to\bar{m}'$, $\w'\to\bar{\w}'$ and $P_{y,L} \to P_{y,R}$. To satisfy the first of these three conditions for general states without over-restricting the spectrum, we must set $a_i=l_i$ and $b_i = r_i$. Indeed, in this case the first condition becomes \eqref{nullgaugeconstrW}, while the last two conditions both reduce to \eqref{eq: null gauge constraints-3}. 
The compensating shifts then take the form  
\begin{equation}
\label{shiftSU2}
  \w' \to \w' - l_2 \, q \ , \quad \ \bar{\w}' \to \bar{\w}' - r_2 \, q ,
\end{equation}
for the left and right $SU(2)$ spectral flow charges, respectively,
\begin{equation}
\label{shiftE}
 E \to E + l_3 \, q =E + r_3 \, q ,  
\end{equation}
for the energy, and
\begin{equation}
\label{shiftY}
n_y \to n_y - \frac{R_y}{2} \left(l_4+r_4\right)q 
 \ , \ \quad
 \w_y \to  
  \w_y - \frac{1}{2 R_y}\left(l_4-r_4
  \right)q ,
\end{equation}
for the $\SSS_y^1$ quantum numbers.

For Eqs.~\eqref{shiftSL2} and \eqref{shiftSU2}--\eqref{shiftY} to make sense in terms of integer spectral flows and momentum/winding numbers, the gauging parameters must be quantised in a specific way.
On the one hand, taking into account that, as argued above, $\w' \pm \bar{\w}'$ must be even, for the $SU(2)$ sector we find $l_2 \pm r_2 \in 2\mathbb{Z}$. We can thus write as a first pass (to be refined momentarily)
\begin{equation}
\label{integersCFT1}
   l_2 = \m + \n \ , \ 
    r_2 = - (\m - \n) \ , \qquad \m,\n \in \mathbb{Z} \,,
\end{equation}
where the signs are chosen for later convenience.
Furthermore, recall that in the $SL(2,\R)$ and $SU(2)$ sectors the spectral flow operations do not act solely on the bosonic sub-algebras. Indeed, they also shift the fermionic modes as in Eqs.~\eqref{flowfermionmodesSL2} and \eqref{flowfermionmodesSU2}. At the level of the vertex operators, this is accounted for by including the $H_{1,2}$ exponentials introduced in  \eqref{flowH12}, which were taken into account for computing the weights \eqref{VirasoroNullWinding}. However, if as in the computation above we start from an unflowed state with no fermionic excitations, and use the shifts \eqref{shiftSL2} and \eqref{shiftSU2} with, say, $q=1$ (or any other odd value), the presence of these exponentials also indicates that the fermion numbers on the left- and right-handed components will \textit{not} be preserved for arbitrary values of $\m \pm \n$. Thus, we see that it is necessary to make Eq.~\eqref{integersCFT1} more precise by restricting to  
\begin{equation}
\label{integersCFT1odd}
   l_2 = \m + \n \in 2\mathbb{Z}+1 
    \ , \qquad
    r_2 = - (\m - \n) \in 2\mathbb{Z}+1 \,, \qquad \m,\n \in \mathbb{Z} \,.
\end{equation}
On the other hand, from \eqref{shiftY} we also must have  
\begin{equation}
\label{integersCFT2}
\frac{1}{2 R_y} \left(r_4-l_4\right) =  \k \in \mathbb{Z}
    \ , \qquad
\frac{R_y}{2} \left(l_4+r_4\right) = \pp \in \mathbb{Z} \,,
\end{equation}
or equivalently 
\begin{equation}
\label{integersCFT2-b}
    l_4\,=\,-\Big(\k R_y - \frac{\pp}{R_y} \Big) \,,\qquad  r_4 \,=\, \k R_y +\frac{\pp}{R_y}\,.
\end{equation}
By plugging the expressions \eqref{integersCFT1odd} and \eqref{integersCFT2-b} into the null constraints \eqref{eq: null gauge constraints}, we now solve for $l_3$, $r_3$ and $\pp$. Firstly, we obtain
\begin{equation}
\label{L32def}
l_3 \;=\; r_3 \;=\; -\sqrt{ 
    \k^2 R_y^2 + \frac{\pp^2}{R_y^2} +   n_5\left(\m^2 + \n^2 - 1\right)},
\end{equation}
where we have chosen the negative square root for $l_3$, $r_3$, so that we gauge away the difference of the upstairs time directions. This fixes
the energy shift \eq{shiftE} in terms of $\k,\m,\n,\pp$. 
More interestingly, we also get  
\begin{equation}
\label{pkequalsn5mn}
    \k \, \pp = n_5\, \m \, \n,
\end{equation}
which shows that only three of the integers $\k,\m,\n,\pp$ are actually independent. Moreover, either $\k$ or $\pp$ (or both) must be even.

We will show below that the integers $\m$ and $\n$ introduced above control the angular momenta of the classical configuration along the $\SSS^3$ coordinates $\phi$ and $\psi$, respectively.
We will further argue in the following that the absolute value of the integer $\k$ is to be interpreted as an orbifold parameter. The meaning of the remaining integer $\pp$ is slightly complicated to interpret in classical terms. This is due to its stringy nature, and it can be understood  either holographically or in terms of T-duality, as follows. 

For $\k\neq 0$, from \eq{pkequalsn5mn} we find that $\pp$ is $n_5$ times the momentum per strand $\m \n/\k$ in the holographic description of JMaRT states, as noted in~\cite{Martinec:2018nco}. This must be an integer since the holographic CFT is a symmetric product orbifold theory (see the discussions in~\cite{Giusto:2012yz,Chakrabarty:2015foa}).  
On the other hand, it is well known that the worldsheet theory is invariant under T-duality along a circular direction. In the language of gauged WZW models, T-dual models arise due to the equivalence of vector and axial gaugings. In the present context, T-duality along $y$ amounts to $l_4\to -l_4$. Together with the usual radius redefinition $R_y \to 1/R_y$, this  exchanges the role of the integers $\k$ and $\pp$. Thus, depending on the choice of duality frame, either $\k$ or $\pp$ are interpreted as an orbifold parameter, while the remaining integer is fixed in terms of $n_5$, $\m$ and $\n$, and it controls the momentum charge.

While on the subject of the T-duality, let us also observe that \eqref{integersCFT2-b} has special features at the self-dual radius, which in $\alpha'=1$ units is at $R_y=1$. For instance, the expressions of the gauging parameters  $l_4,r_4$ corresponding to the $U(1)_y$ component resemble those of their $SU(2)$ counterparts $l_2,r_2$. 
In addition, we see that it becomes possible to set either $l_4$ or $r_4$ to zero while keeping the other one non-trivial, which is not possible for generic values of $R_y$. Since the self-dual radius is associated with the appearance of new massless states (in the upstairs theory), this might lead to new solutions. We leave a more detailed exploration of such configurations for future work.

Note that the values of the gauging parameters \eqref{integersCFT1odd}--\eqref{L32def} imply that
\begin{align}
\label{SigmapostiveCFT}
\begin{aligned}
    \Sigma_0\left(\rho=0,\theta=0\right) \;=\; \frac{1 - l_2 r_2}{2} + \frac{l_3r_3 - l_4r_4}{n_5} &\;=\; 
     \m^2 + \frac{\k^2 R_y^2}{n_5} \,,\\
    \Sigma_0\left(\rho=0,\theta=\frac{\pi}{2}\right) \;=\; \frac{1 + l_2 r_2}{2} + \frac{l_3r_3 - l_4r_4}{n_5}  &\;=\; \n^2 + \frac{\k^2 R_y^2}{n_5}  \, .
\end{aligned}
\end{align}
The combinations in \eqref{SigmapostiveCFT} correspond to the minimal values of the quantity $\Sigma_0$ appearing in the denominator of various components of the supergravity fields~\eqref{eq:metric-B-gen}. We have just shown that the consistency conditions of the spectrum imply that they are both non-negative quantities. This will be important when studying the corresponding geometry in the following section.

\medskip

Let us summarise our results so far. Starting with a class of generic null gauged models, the gauged currents are defined in terms of eight parameters, namely $\lL_i,\rR_i$, $i=1,2,3,4$. Since only the direction of the gauging matters, the overall scale becomes irrelevant. This means that we can work directly with the six ratios $l_i, r_i$. These must satisfy the two null conditions \eqref{eq: null gauge constraints-3} and also $l_3=r_3$ from the non-compactness of $t$. Finally, by focusing on the worldsheet CFT and relating the action of spectral flow in $SL(2,\R)$ to that of the gauge orbits, we have shown that the theory is consistent only if the remaining three parameters can be written in terms of three integers (in addition to $R_y$), which we can take to be  $\m, \n$ and $\k$ (when $\k \neq 0$). Moreover, these must be chosen so that $\pp= n_5 \m \n/\k$ is also integer-valued. It is possible that the quantization conditions on $\k,\m,\n,\pp$ could alternatively be obtained by analyzing the global consistency of the gauging, see e.g.~\cite{Gawedzki:1991yu,Hull:2006qs,Hull:2006va,Gawedzki:2010rn,deFromont:2013iy}. In the next section we will instead proceed to analyze the global geometry of the gauged target space.

The set of conditions \eqref{integersCFT1odd}--\eqref{L32def} is one of the main results of this paper. It will allow us to rewrite the general supergravity fields in Eqs.~\eqref{eq:metric-B-gen}--\eqref{eq:dilaton-pre} in a simple way, making their main physical features and some of their symmetries manifest. Furthermore, this will lead to a complete characterisation of the full set of consistent solutions.


\section{Analysis of the supergravity backgrounds} 
\label{sec:gen-CTC}

In the previous section we have shown that the gauging parameters $l_i,r_i$, $i=1,2,3$ can be defined in terms of  $\k,\m,\n,\pp$, all of which are integers that have a clear physics meaning. 
We now perform an independent supergravity analysis of the metrics introduced in Eq.~\eqref{eq:metric-B-gen}, and show that imposing smoothness and absence of closed timelike curves provides an alternative derivation and complementary interpretation of the constraints \eqref{integersCFT1odd}--\eqref{L32def}.

\subsection{Eliminating potential closed timelike curves}

To investigate potential closed timelike curves we complete the squares successively in the periodic variables $\psi$, $\phi$ and  $y$, to rewrite the line element \eqref{eq:metric-B-gen}. We obtain
\begin{align}
	ds^2 = & - T(\rho) dt^2 + Y(\rho) \left[  dy + A_y (\rho ; dt)\right]^2 + n_5 (d\rho^2 +  d\theta^2 ) + \nonumber\\
	& + n_5 \,\sin^2 \theta  \, \frac{h_\phi}{\Sigma_0} \left[  d\phi + A_\phi (\rho ; dt,dy)\right]^2 +  n_5 \,\cos^2 \theta  \, \frac{h_\psi}{\Sigma_0} \left[  d\psi + A_\psi (\rho ; dt,dy)\right]^2 ,
\end{align}
where the functions $\Sigma_0$, $h_\phi$ and $h_\psi$  were defined in \eqref{eq:Sigma} and \eqref{eq:hs} respectively, $ A_y$, $A_\psi$, and  $A_\phi $ are one-forms depending only on the radial variable $\rho$ and with legs in the appropriate arguments, while  $T(\rho)$ will turn out to be a non-negative function whose explicit expression we will not need. 
In order to ensure the absence of CTCs we must require the functions multiplying the squares in the periodic variables to be non-negative, i.e.
\begin{align}
\label{eq:ineqCTC}
    \frac{h_\phi}{\Sigma_0}  \geq 0 \, , \qquad
    \frac{h_\psi}{\Sigma_0}  \geq 0 \, , \qquad
    Y(\rho)  \geq 0.
\end{align} 
We now show that asking for the inequalities \eqref{eq:ineqCTC} to hold everywhere in geometry is equivalent to imposing 
\begin{equation}
    \label{eq: No CTC ratio}
    l_3 = r_3. 
\end{equation}
Let us first see that \eqref{eq:ineqCTC} implies \eqref{eq: No CTC ratio}. By combining the first two inequalities we find that the product $h_\phi h_\psi$ is non-negative. On the other hand, the explicit expression of $Y(\rho; l_i,r_i)$ reads
\begin{align}
    Y(\rho) &\,=\, 
    \frac{4 \sinh^2\rho\left(n_5\cosh^2\rho+l_3r_3\right)
    + (n_5+l_3r_3)^2 - (n_5 l_2^2+l_4^2)(n_5 r_2^2+r_4^2)}{4 n_5^2 h_{\phi}h_\psi} \nn \\
    &\,=\,
    \frac{4n_5\sinh^2\rho\left(n_5\cosh^2\rho+l_3r_3\right)-\left(l_3-r_3\right)^2}{4 n_5 h_{\phi}h_\psi}\,, \label{eq:explicitY}
\end{align}
where in the second line we have used \eqref{eq: null gauge constraints-3}. It follows from this last expression that the third inequality in \eqref{eq:ineqCTC} can only be satisfied at the origin $\rho=0$ if $l_3 = r_3$. 

It remains to be seen that the implication holds in the other direction as well.  For this, we note that the minimal value of $\Sigma_0$ is given by 
\begin{equation}
    \Sigma_0^{\mathrm{min}} \,=\, \frac{1}{2n_5} \big[n_5(1-|l_2 r_2|) + l_3 r_3 - l_4 r_4\big].
\end{equation}
Using $l_3 = r_3$ we can rewrite the null conditions \eqref{eq: null gauge constraints-3} as
\begin{equation}
    n_5 \left(l_2^2 - r_2^2\right) + r_4^2 - l_4^2 \;=\; 0 \ , \ \quad~~ 
    l_3^2 + r_3^2 \;=\; 2 l_3^2 \;=\; l_4^2 + r_4^2 + n_5 \left(
    l_2^2 + r_2^2 -2 \right), 
\end{equation}
so that
\begin{equation}
    2 \big[n_5 (1 \pm l_2r_2) + l_3 r_3 - l_4 r_4\big] \;=\;
    n_5 (l_2 \pm r_2)^2 + (l_4 - r_4)^2 \,. 
\end{equation}
It follows that $\Sigma_0\ge 0$ everywhere. Given that $h_\phi = \Sigma_0(\rho,\theta=0)$ and $h_\psi = \Sigma_0(\rho,\theta=\pi/2)$, the same holds for these functions and the first two inequalities in \eqref{eq:ineqCTC} are thus satisfied. The third one also holds, as can be checked from \eqref{eq:explicitY}.

This proves that in the asymptotically linear dilaton  geometry the necessary and sufficient condition for avoiding CTCs is precisely $l_3 = r_3$, Eq.~\eqref{eq: No CTC ratio}. This constraint was also obtained in the worldsheet analysis of Section \ref{sec:CFTanalysis} from the non-compactness of the $t$ direction, Eq.~\eqref{eq:l3equalr3}. Moreover, once this is imposed we find that, as advertised above, $T(\rho)$ is non-negative.

\subsection{Absence of horizons}

Here and in the following subsection we perform an analysis which closely follows that of  \cite{Jejjala:2005yu}.
The determinant of the metric \eqref{eq:metric-B-gen} reads 
\begin{equation}
\label{detg}
    \det \:\! g = - \left( \frac{n_5^2 \sin(2 \theta) \sinh(2 \rho)}{4 \Sigma_0}   \right)^2,
\end{equation}
where we have used the null-gauge constraints \eqref{eq: null gauge constraints-3}.
Besides the usual zeros at the poles of the $\SSS^3$, this only vanishes at $\rho=0$. Given that the determinant of the induced metric on surfaces of constant $\rho$ is simply \eqref{detg} divided by $n_5$,  we see that $\rho=0$ corresponds to either a horizon or an origin of higher codimension.  
In order to distinguish between these two cases, we further compute the determinant of the induced metric on surfaces of constant $\rho$ and $t$, and we evaluate it at $\rho=0$,  giving %
\begin{equation}
\label{detgsub}
    \lim_{\rho \to 0} \det \:\! g\big|_{(y,\theta,\phi,\psi)}  = - \left( \frac{n_5(l_3-r_3) \, \sin(2 \theta)}{4 \Sigma_0(0,\theta)}   \right)^2 \,.
\end{equation}
Hence, in order to obtain a horizonless and possibly smooth geometry we again need to impose $l_3=r_3$, Eq.~\eqref{eq: No CTC ratio}, such that \eqref{detgsub} vanishes. Smoothness will then be achieved if some circle direction shrinks appropriately when $\rho \to 0$, as we discuss next.  

\subsection{Smoothness and quantisation}

When the background contains F1 charge, the supergravity solutions must be smooth up to possible orbifold singularities. In the absence of F1 charge, NS5-brane singularities will be present. We begin by treating the more general case in which F1 charge is present, and treat the latter as a special case.

We therefore focus on the periodic directions and consider a generic Killing vector of the form
\begin{equation}
\label{eq:xismoothness}
    \xi = \del_y + \alpha \, \del_\psi - \beta \, \del_\phi \; , \qquad \alpha, \beta \in \RR.
\end{equation}
where the signs have been chosen for later convenience.
To find smooth solutions we seek pairs of coefficients $(\alpha,\beta)$ such that the norm of $\xi$ vanishes $\forall \, \theta \in [0, \frac{\pi}{2}] $ when we approach $\rho =0$. From the metric \eqref{eq:metric-B-gen} we find that this is indeed the case when
\begin{equation}
\label{eq: alpha beta fixed}
    \alpha = \frac{l_2 r_4 + l_4 r_2}{2 n_5 \Sigma_0(0,\frac{\pi}{2})} \; , \qquad     \beta = \frac{l_2 r_4 - l_4 r_2}{2 n_5 \Sigma_0(0,0)} \,,
\end{equation}
since for these values, and upon using the null constraint \eqref{nullgaugeconstr}, we obtain
\begin{equation}
\lim_{\rho \to 0} g_{ij} \, \xi^i \, \xi^j =  -  \frac{(l_3 - r_3 )^2}{4 n_5 \Sigma_0(0,0)\; \Sigma_0(0,\frac{\pi}{2}) } = 0 \; ,
\end{equation}
where $i,j = y, \psi, \phi$. In the last step we have used the no-CTC condition \eqref{eq: No CTC ratio}. 
Then, we define the following shifted coordinates
\begin{equation}
    \hat{\psi} = \psi + \alpha \, y \; , \qquad \hat{\phi} = \phi - \beta \, y \;,
\end{equation}
where the signs are chosen for later convenience. By examining the integral curves of $\xi$, we see that the direction that shrinks at $\rho=0$ is $y$ at fixed $\hat{\psi}$, $\hat{\phi}$.
We find that near $\rho=0$ the line element at fixed $(t,\theta,\hat{\psi},\hat{\phi})$ is of the form
\begin{equation}
    \label{eq:metrickRy}
    ds^2_{\rho \to 0} \simeq n_5  \left[d\rho^2 + \rho^2 \, d\left(\frac{y}{R}\right)^2  \right] \, , \qquad 
     R^2 = \left[\frac{2 n_5^2 \Sigma_0(0,0) \, \Sigma_0(0,\frac{\pi}{2})}{ \left( n_5 + l_3^2 \right) (l_4 - r_4)} \right]^2, 
\end{equation}
where we have used \eqref{eq: null gauge constraints} and \eqref{eq: No CTC ratio}, and assumed $l_4 \neq r_4$ (for now).
Strictly speaking, a smooth geometry will be obtained only if the radius $R$ coincides with $R_y$. Given that string theory is well-defined on orbifold backgrounds, as usual we allow for possible $\mathbb{Z}_\k$ orbifold singularities, $\k$ being the corresponding orbifold parameter. Thus, we relax this condition and impose $R^2=\k^2 R_y^2$ for some positive integer $\k$ instead, i.e.~ 
\begin{equation}
    \label{eq: Radius fixed}
    \k^2 R_y^2 = \left[\frac{2 n_5^2 \Sigma_0(0,0) \, \Sigma_0(0,\frac{\pi}{2})}{ \left( n_5 + l_3^2 \right) (l_4 - r_4)} \right]^2.
\end{equation}
Making use of the intuition developed in Section \ref{sec:CFTanalysis},  we further rewrite the values of the parameters $l_i, r_i$ in terms of new quantities $\m$, $\n$, $\k$ and $\pp$ as in Eqs.~\eqref{integersCFT1} and \eqref{integersCFT2}, namely 
\begin{align}
\label{eq: new parametrisation}
    l_2 = \m + \n \ , \  r_2 = -(\m - \n )
    \ , \ 
    \half \left( l_4 - r_4 \right)  = - \k R_y 
    \ , \ 
    \half \left( l_4 + r_4  \right) =  \frac{\mathsf{p}}{R_y} =  \frac{n_5}{R_y} \frac{\m \n }{\k}
    .
\end{align}
In \eq{eq: new parametrisation} we could a priori have written $\k'$ instead of $\k$. However, if we were to then substitute \eq{eq: new parametrisation} into \eq{eq: Radius fixed}, we would find that $\k'=\pm \k$. In other words, within this parametrisation Eq.~\eqref{eq: Radius fixed} is trivially satisfied. 

Up to this point, the reparametrisation \eqref{eq: new parametrisation} does not assume that $\m$ and $\n$ are integers. However, the periodicities of the new angular variables $\hat{\phi}$ and $\hat{\psi}$ should be consistent with that of $y$. The corresponding quantisation conditions read
\begin{equation}
\label{quantisation geometry orbifold}
\alpha \, (\k\, R_y) = \m \in \ZZ \;, \qquad \beta \, (\k \, R_y) = \n \in \ZZ \,. 
\end{equation}
From the classical point of view, the values of the integers $\m$ and $\n$ seem otherwise unrestricted. However, we know from the discussion around Eq.\;\eqref{integersCFT1odd} that one of $\m$, $\n$ must be even and the other one must be odd. A geometric argument leading to this restriction was put forward in \cite{Jejjala:2005yu} in the JMaRT context. This is based  on discussing the periodicity of the fermions along the $\SSS^1_y$ circle and the associated spin structure of the target space. Although it should be possible, we will not attempt to extend these arguments to the case with general gauging parameters. This is because in the next section we will directly match the JMaRT solutions to the supergravity fields of our coset models.

Moreover, out of the four parameters $\m,\n,\k$ and $\mathsf{p}= n_5 \m \n / \k$, a priori only the first three appear to be required to be integers from the above smoothness analysis, and nothing seems to prevent $\mathsf{p}$ from being a rational (not necessarily integer) number. The stringy nature of this parameter manifests itself in the fact that T-duality along $\SSS^1_y$, namely $l_4 \rightarrow - l_4$ and $R_y \rightarrow 1/R_y$, maps $\k \leftrightarrow  \mathsf{p}$. Since in the T-dual geometries $\pp$ is an orbifold parameter, it must also be quantised. Moreover, given that $\m$ and $\n$ may vanish or be non-vanishing integers with arbitrary signs, so can $\pp$, and consequently, the same applies to $\k$. In due course we will restrict to non-negative values of $\k,\m,\n,\pp$, without loss of generality.

Recall that in the previous passage we assumed $l_4 \neq r_4$. The case $l_4-r_4=0$, which corresponds to $\k=0$ in the parametrisation \eqref{eq: new parametrisation}, needs to be treated separately. 
When $\k=0$, $\Sigma_0$ goes to zero at $\rho=0$ and either $\theta=0$ or $\theta=\pi/2$, see e.g.~Eq.\;\eqref{SigmapostiveCFT}.
The metric is singular as  $\Sigma_0 \to 0$. We choose conventions in which the zero is at $\theta=\pi/2$. This corresponds to the location of the (smeared) NS5 brane source. As we shall see in the next section, this is because the F1 charge vanishes and the solutions are two-charge NS5-P (see Eqs.~\eqref{Q1Qpb2first} and~\eqref{MetricandBnewNS5P}--\eqref{defsnewNS5P} below).
Let us analyse the geometry away from the source. The region of interest is the neighbourhood of $\rho=0$ for $\theta\neq\pi/2$. Note that, assuming $l_3 = r_3$, the null conditions \eqref{eq: null gauge constraints-3} imply $l_2 = \pm r_2$, so that either $\m=0$ or $\n=0$. To have the source at $\theta\neq\pi/2$, we take $l_2=-r_2$, i.e.~$\n=0$. 
Then the norm of the Killing vector \eqref{eq:xismoothness} is always non-vanishing. 
However, the $\psi$ circle shrinks as $\rho \to 0$. Indeed, in this neighbourhood the line element at fixed $(t,y,\theta,\phi)$ reads
\begin{equation}
\label{eq:m-orb}
    ds^2_{\rho \to 0} \simeq n_5  \left[d\rho^2 + \rho^2 \, d\left(\frac{\psi}{\m}\right)^2  \right] \, ,
\end{equation}
where we have used the parametrisation \eqref{eq: new parametrisation} without imposing the last equality, so that $\pp$ is unconstrained. We conclude that for generic values of $\theta$ and $\m=\pm 1$ the geometry is smooth. As before, we allow for orbifold singularities, so that $\m$ is again quantised: any other non-zero $\m \in \mathbb{Z} $ leads to a $\mathbb{Z}_{|\m|}$ orbifold structure. Had we chosen $l_2 = r_2$ instead, the source would have been at $\theta=0$, the $\phi$ circle would have been the one shrinking at small $\rho$, and the parameter $\m$ would have been replaced by $\n$.

\subsection{Killing Spinors}

Finally, we discuss the relation that the embedding coefficient must satisfy in order to preserve a certain amount of supersymmetry. For simplicity, we work directly in the $AdS_3\times \SSS^3$ limit, which will be shown to exist in all consistent cases. In order to achieve this, we first consider the Killing spinor in global $AdS_3$, as given in \cite{Jejjala:2005yu}:
\begin{equation}
    \epsilon^\pm_L = e^{\pm \frac{i}{2} \tilde{\phi}_L} \, e^{- \frac{i}{2} y} \epsilon_0 \, , \qquad \epsilon^\pm_R = e^{\pm \frac{i}{2} \tilde{\phi}_R} \, e^{- \frac{i}{2} y} \epsilon_0\, ,
\end{equation}
where $\epsilon_0$ is a constant $AdS_3$ spinor. 
The dependence on the spacetime coordinates was derived in \cite{Izquierdo_1995} and, in particular, the $y$ dependence is such that the Killing spinors are regular near the origin (see also \cite{Balasubramanian:2000rt} and \cite[App.~D,E]{Mandal_2008}). By a large gauge transformation one can induce the $\SSS^3$ angular momenta starting from global $AdS_3 \times \SSS^3$ which, in terms of the $\tilde{\phi}_{L,R}$ coordinates, translates into the following diffeomorphism
\begin{equation}
    \tilde{\phi}_L = \phi + (\m + \n) \, y  = \, \phi + l_2 \, y \; , \qquad \tilde{\phi}_R = \phi + (\m - \n) \, y \, = \phi - r_2\, y \,.
\end{equation}
This is known as spacetime spectral flow (see e.g.~\cite{Giusto:2012yz,Chakrabarty:2015foa}).
Focusing on solutions that admit an asymptotically flat completion, the Killing spinor equations  demand that the spinors be independent of $y$ after performing the above large gauge transformation. This is obtained by imposing 
\begin{equation}
    |l_2| = 1 \ \quad \mathrm{or} \ \quad |r_2| = 1. 
\end{equation}
By virtue of the null condition \eqref{eq: null gauge constraints}, the above constraint implies an analogous one for the remaining embedding coefficients, namely 
\begin{equation}
    |l_3| = |l_4| \ \quad \mathrm{or} \ \quad |r_3| = |r_4| \,. 
\end{equation}
We thus conclude that the above constraints must be satisfied in order to have non-trivial Killing spinors in spacetime. This is consistent with the discussion around Eqs.~\eqref{susyrestricparam} and \eqref{susyrestricparamR} above. 

\vspace{0.4cm}

Summarising, we have imposed absence of CTCs, absence of horizons, and smoothness up to physical sources (corresponding to orbifold singularities or NS5 branes) of the general background \eqref{eq:metric-B-gen}. These conditions imply a set of constraints on the group-theoretic embedding coefficients $l_i,r_i$ parametrising the space of solutions in which the string propagates without pathologies. These consistency conditions take exactly the same form as those obtained from the worldsheet CFT analysis in Section~\ref{sec:CFTanalysis}, Eqs.~\eqref{integersCFT1odd}--\eqref{L32def}.  
In passing, we note that recently a similar relation between consistency of the worldsheet theory and a well-behaved geometry was found in a related context in \cite{Chakraborty:2018vja,Chakraborty:2019mdf}.


\section{Matching to JMaRT and two-charge limits} 
\label{sec:JMaRTuniqueness}

In the previous sections we have shown that the class of null-gauged models defined 
in Section~\ref{sec:review} in terms of the gauging parameters $l_i,r_i$ are consistent 
iff the latter can be written simply in terms of the four integers $\k,\m,\n$ and $\pp$. 
Out of these, only three are independent. Here we show that the resulting models correspond precisely to the full family of JMaRT solutions \cite{Jejjala:2005yu} and their various limits. The  metric and $B$-field take a simple form in terms of these parameters, all of which have a clear physical meaning. Moreover, starting from the generic three-charge solution, we describe in detail the delicate limits that lead to the two-charge configurations and exhibit novel non-BPS NS5-P solutions.

\subsection{JMaRT metric and $B$-field}

Let us recall the form of the NS5-decoupled limit of the NS5-F1-P JMaRT solutions, that is, the $S$-duals of the smooth and horizonless non-supersymmetric D1-D5-P backgrounds obtained in \cite{Jejjala:2005yu}. Note that we are working in units where $\alpha' = 1$, hence $Q_5 = n_5$. In our conventions, these geometries take the form \cite{Martinec:2018nco}
\begin{align}
\begin{aligned}
\label{eq:JM-orig}
	ds^2 \;=&\; \frac{f}{\tilde{H}_1} \left(-dt^2 + dy^2 \right) + \frac{M}{\tilde{H}_1} \left( c_p dt - s_p dy \right)^2 + n_5 \left(d\rho^2 + d\theta^2  \right)   \\
	&+ \frac{n_5 }{\tilde{H}_1} \Big[ \left(r_+^2 - r_-^2\right) \cosh^2 \rho + r_-^2 + a_2^2 + Ms_1^2   \Big] \, \sin^2 \theta \, d\phi^2   \\
	&+ \frac{n_5 }{\tilde{H}_1} \Big[ \left(r_+^2 - r_-^2\right) \sinh^2 \rho + r_+^2 + a_1^2 + Ms_1^2   \Big] \, \cos^2 \theta \, d\psi^2   \\
	& + \frac{2\sqrt{M \, n_5}}{\tilde{H}_1} \Big[ \left( a_2c_1c_p - a_1 s_1 s_p  \right) dt + \left( a_1 s_1 c_p - a_2 c_1 s_p   \right) dy  \Big]\, \sin^2 \theta \, d\phi   \\
	& + \frac{2\sqrt{M \, n_5}}{\tilde{H}_1} \Big[  \left( a_1c_1c_p - a_2 s_1 s_p \right) dt + \left( a_2s_1c_p - a_1c_1s_p \right) dy   \Big]\, \cos^2 \theta \, d\psi   \\
	&+ ds^2_{\TT^4} \, ,
\end{aligned}
\end{align}
\begin{align}
\begin{aligned}
\label{eq:JM-orig-B}
B\; =&\; - \frac{M s_1 c_1}{\tilde{H}_1} \, dt \wedge dy + \frac{n_5 \cos^2 \theta }{\tilde{H}_1} \Big[ (r_+^2 - r_-^2)\sinh^2 \rho + r_+^2 + a_2^2 + Ms_1^2  \Big] d\phi \wedge d\psi \\
	&{}+ \frac{\sqrt{M n_5}}{\tilde{H}_1} \Big[ (a_1 c_1 c_p - a_2 s_1 s_p) dt + (a_2 s_1 c_p - a_1 c_1 s_p )dy  \Big] \wedge  \sin^2 \theta \,  d\phi \\
	&{}+ \frac{\sqrt{M n_5}}{\tilde{H}_1} \Big[ (  a_2 c_1 c_p - a_1 s_1 s_p ) dt + ( a_1 s_1 c_p -  a_2 c_1 s_p ) dy  \Big] \wedge  \cos^2 \theta \,  d\psi, 
\end{aligned}
\end{align}
together with the dilaton 
\begin{equation}
\label{DilatonJMART}
    e^{2 \Phi} = g_s^2 \,\frac{n_5}{\tilde{H}_1}.
\end{equation}
Here the charges are given in terms of the boost parameters $\delta_{1,p}$, that is 
\begin{equation}
Q_i = M s_i c_i \ , \ c_i = \cosh(\delta_i) \ , \ s_i = \sinh(\delta_i)  \qqquad i = 1,p \; , 
\end{equation}
and
\begin{align}
&f = \half \left[ (r_+^2 - r_-^2) \cosh(2\rho) + (a_2^2 - a_1^2) \cos(2\theta) + r_+^2 + r_-^2 + a_1^2 + a_2^2  \right] \, , \nn \\
&\tilde{H}_1 = f + M s_1^2 = \half \left[ (r_+^2 - r_-^2) \cosh(2\rho) + (a_2^2 - a_1^2) \cos(2\theta) +M^2  \right] + M s_1^2 \, , \\
&r_\pm^2 = \half \Big[ (M-a_1^2 - a_2^2) \pm \sqrt{(M-a_1^2 - a_2^2)^2 - 4a_1^2 a_2^2}   \Big] = - a_1 a_2 \left(\frac{s_1s_p}{c_1c_p}\right)^{\pm 1}
\, . \nn
\end{align}
In these formulas, parameters such as $a_{1,2}$ are a priori thought of as continuous. As will be clear shortly, this is potentially misleading since they are constrained by the smoothness conditions and the absence of horizons. In this setup, these constraints read
\begin{align}
\begin{aligned}
a_1 a_2 &\;=\; \frac{Q_1 Q_5}{\k^2 R_y^2} \frac{s_1^2 c_1^2 s_p c_p}{(c_1^2 c_p^2-s_1^2 s_p^2)},\cr
M &\;=\;  a_1^2 + a_2^2+r_+^2 + r_-^2 \;=\;  a_1^2 + a_2^2 - a_1 a_2 \frac{c_1^2 c_p^2 + s_1^2 s_p^2}{c_1 c_p s_1 s_p} \,,
\end{aligned}
\end{align}
and
\begin{equation}
\m  = \sqrt{\frac{M}{n_5}} \frac{\k R_y s_p c_p}{
(a_2 s_1 s_p-a_1 c_1 c_p)} \ \ \in \mathbb{Z} \ , \quad~~
\n = \sqrt{\frac{M}{n_5}} \frac{\k R_y s_p c_p}{
(a_2 c_1 c_p-a_1 s_1 s_p)} \ \ \in \mathbb{Z}.
\end{equation}
Here the integer numbers $\m$ and $\n$ again parametrise the angular momenta on $\SSS^3$. Moreover, these constraints imply the important identity 
\begin{equation}
\label{Q1Qpratio}
    \frac{Q_p}{Q_1} \;=\; \frac{n_5 \, \m\, \n}{\k^2 R_y^2} \;=\; \frac{\pp}{\k} \frac{1}{R_y^2}
\end{equation}
with $\k \, \pp = n_5 \, \m \, \n$ as before, which relates non-trivially the three charges sourcing the configuration.   
We now show that the above set of conditions leads to a further set of relations which enable us to rewrite the JMaRT solutions in terms of the three integers $\k$, $\m$ and $\n$, together with a single dimensionful scale set by $R_y$. A related but different calculation was carried out in \cite{Gimon:2007ps}.
Defining
\begin{equation}
    b^2 = r_+^2 - r_-^2 \ \ \Rightarrow \ \ f = b^2 f_0 \ \  \Rightarrow \ \ \tilde{H_1} = b^2 \Sigma_0 \,, 
    \label{H1tb2}
\end{equation}
the most useful relations are of the following form: 
\begin{align}
a_2^2 - a_1^2   = b^2( \m^2 - \n^2) \ , \ 
f + M \, s_p^2  & = b^2 \, h_y \ , \
f - M \, c_p^2 = b^2 \, h_t \, ,  \\
M (c_1^2 + s_1^2)  = b^2 \left( \m^2 + \n^2 -1 + \frac{2\k^2 R_y^2}{n_5}    \right) \ & , \ 
M (c_p^2 + s_p^2) = b^2 \left( \m^2 + \n^2 -1 + \frac{2\pp^2 }{n_5 R_y^2}    \right) \, , \nn \\ 
a_1^2 + r_+^2 + M s_1^2   = b^2 \left( \n^2 + \frac{\k^2 R_y^2}{n_5} \right) \ &, \
a_2^2 + r_+^2 + M s_1^2   = b^2 \left( \m^2 + \frac{\k^2 R_y^2}{n_5} \right) \nn \, , \\
\sqrt{M \, n_5} \left( a_2 c_1 s_p - a_1 s_1 c_p   \right)  = b^2 \left( \m \frac{\pp}{R_y} + \n \, \k R_y  \right) \ &, \
\sqrt{M \, n_5} \left( a_2 s_1 s_p - a_1 c_1 c_p   \right)  = b^2 \, \n \, \Delta \nn \, , \\
\sqrt{M \, n_5} \left( a_2 s_1 c_p  - a_1 c_1 s_p   \right) = b^2 \left( \n \frac{\pp}{R_y} + \m \, \k R_y  \right) \ &, \ \sqrt{M \, n_5} \left( a_2 c_1 c_p - a_1 s_1 s_p   \right) = b^2 \, \m \, \Delta
 \, , \nn
\end{align}
where we have defined 
\begin{align}
    \Sigma_0& = 
    \sinh^2\rho + (\m^2-\n^2) \cos^2\theta +  \n^2 + \frac{\k^2 R_y^2}{n_5} \, ,  
    \label{Sigma0def} \\
    \Delta & = \sqrt{n_5(\m^2+\n^2-1)+ \k^2 R_y^2 + \frac{\pp^2}{R_y^2}} \, , 
    \label{Deltadef} 
\end{align}
such that $\Sigma_0$ is the same quantity as in previous sections.
Finally, we have 
\begin{equation}
\label{Q1Qpb2first}  
\frac{Q_1}{b^2}  = \, \frac{\k R_y}{n_5} \, \Delta  \ , \ \quad
\frac{Q_p}{b^2}  = \, \frac{\pp }{n_5R_y } \, \Delta \,. 
\end{equation}
We note that for the metric and the $B$-field we do not need the individual charges $Q_1$ and $Q_p$, but only the ratios \eqref{Q1Qpb2first}. By using these formulas, the $b^2$ factor cancels out completely, and we finally obtain the six-dimensional fields
\begin{align}
\label{FinalMetricandBfield}
      ds^2  \,  =&   \,\;  n_5(d\theta^2 + d\rho^2) + \frac{1}{\Sigma_0}\Bigg[  - \left(\sinh^2\rhoo + (\m^2 - \n^2) \cos^2 \theta + 1 - \m^2 - \frac{\pp^2}{n_5 R_y^2}  \right)   dt^2   \\
    & + \left(\sinh^2\rhoo + (\m^2 - \n^2) \cos^2 \theta + \n^2  + \frac{\pp^2}{n_5 R_y^2}  \right)   dy^2 - 2 \:\! \frac{\pp }{n_5 R_y} \;\! \Delta\;\! dt dy \nn \\[1.5mm]
    & + \left( n_5 \sinh^2 \rhoo + n_5 \m^2 + \k^2 R_y^2 \right)
    \:\! \sin^2 \theta \;\! d\phi^2 + \left( n_5 \sinh^2 \rhoo + n_5 \n^2 + \k^2 R_y^2 \right) \;\! \cos^2 \theta \:\! d\psi^2  \nonumber \\[1.5mm]
    & + 2  \left( \m \:\! \Delta \:\! dt  - \Big(  \m \frac{\pp}{R_y} + \n \:\! \k R_y \Big)  dy \right)  \sin^2 \theta \:\! d\phi - 2 \left( \n  \:\! \Delta \:\! dt  - \Big( \n  \frac{\pp}{R_y} + \m \:\! \k R_y \Big)  dy \right)  \cos^2 \theta \;\! d\psi \Bigg] , \nn \\[3mm]
 B  \:  =&   \,\; \frac{1}{\Sigma_0} \Bigg[  - \frac{\k R_y}{n_5} \;\! \Delta \:\! dt \wedge dy +  \left( n_5 \sinh^2\rhoo + n_5 \,  \m^2 + \k^2 R_y^2  \right) \, \cos^2 \!\theta \,  d\phi \wedge d\psi  
 \nonumber\\
 & + \left( \m \:\!\Delta \:\! dt  - \Big(  \m \frac{\pp}{R_y} + \n \:\! \k R_y \Big) dy \right) \wedge  \cos^2 \theta \:\! d\psi
 -\left( \n \:\! \Delta \:\! dt - \Big(\n  \frac{\pp}{R_y} + \m \:\! \k R_y\Big)  dy \right) \wedge  \sin^2\theta  \:\! d\phi  \Bigg] \:\! . \nn 
\end{align}
This is exactly the geometry we get from the null-gauge construction studied in the previous sections when inserting the parametrisation \eqref{eq: new parametrisation} for the $l_i,r_i$ gauging parameters in Eqs.~\eqref{eq:metric-B-gen}. 

It is worth discussing some interesting facts about the expressions we have presented in \eqref{FinalMetricandBfield}. First, we note the trivial symmetry associated to exchanging the two $\SSS^3$ angular momenta. This corresponds to the re-labelling $\m \leftrightarrow \n$ and $\phi \leftrightarrow - \psi$, which must be accompanied by the shift $\theta \to \pi/2 - \theta$. On the other hand, we note that while the usual JMaRT geometry is obtained by replacing $\pp=n_5 \m \n / \k$ in the expressions in \eqref{FinalMetricandBfield}, here we have chosen a slightly more general form by keeping $\pp$ explicit. As a result, we easily find a symmetry that corresponds to exchanging $\k \leftrightarrow \pp$ and $R_y \to 1/R_y$, which we have identified above as T-duality. At the classical level, we can now see that this operation is equivalent the well-known Buscher rules \cite{Buscher:1987sk}, where $g_{yy} \to 1/g_{yy}$, $g_{ty} \to B_{ty}/g_{yy}$, etc. 

Importantly, by keeping $\pp$ explicit in  \eqref{FinalMetricandBfield} we have presented expressions that are valid even for solutions where $\k=0$. As will be reviewed below, this includes the limit associated to the BPS and non-BPS two-charge NS5-P configurations. 

\subsection{The dilaton }

As described above, the JMaRT dilaton is of the form \eqref{DilatonJMART}. The only coordinate-dependent part of this expression corresponds to $\Sigma_0$ as defined in \eqref{Sigma0def}, where we have used \eqref{H1tb2}. This matches exactly with the expression obtained in Section~\ref{sec:review}, see Eq.~\eqref{eq:dilaton-pre}, by considering the supergravity equations of motion, which provide the dilaton up to a multiplicative constant. The matching with the JMaRT backgrounds thus gives a criteria for choosing this constant appropriately: it is given by $n_5/b^2$, i.e.
\begin{equation}
    e^{2\Phi} = \frac{n_5}{b^2 \Sigma_0}. 
\end{equation}

In order to make the expression for the dilaton more transparent we proceed as follows. First, we introduce the canonical expressions for the charges 
\begin{equation}
    Q_1 = n_1 \frac{g_s^2 }{V_4} \ , \quad \ Q_p = \frac{n_p}{R_y^2} \frac{g_s^2 }{V_4} \, ,
\end{equation}
where $V_4$ is the volume of the internal $\TT^4$, while $n_{1}$ is the number of fundamental string sources and $n_p$ is the integer momentum charge. We observe that the key property \eqref{Q1Qpratio} is equivalent to 
\begin{equation}
\label{kn1pnp}
    \frac{\k R_y}{Q_1} = \frac{\pp}{Q_p R_y}.
\end{equation} 
This ties in nicely with the fact that, as discussed above, T-duality interchanges $Q_1\leftrightarrow Q_p$ and $\k R_y\leftrightarrow \pp/R_y$. It also justifies referring to $\pp$ as being related to the momentum charge as we did in previous sections. We have seen that the parameters $\m$ and $\n$ are associated to the angular momenta of the geometry. Here we find in \eqref{kn1pnp} that $\pp$ and $\k$ relate to the momentum and F1 winding charges of the black hole microstate in question along the asymptotic $y$-circle.

Furthermore, while the $b^2$ factor is irrelevant for writing down the metric and $B$-field, it does appear when computing the dilaton. This means that we need to work with the individual charges $Q_1$ and $Q_p$ as opposed to the ratios. By making use of Eqs.~\eqref{Q1Qpb2first} we obtain two equivalent expressions for the dilaton, namely 
\begin{equation}
\label{FinalDilaton}
    e^{2\Phi} \;=\; \frac{\Delta}{\Sigma_0} \, \frac{\k R_y  }{Q_1} \;=\;
    \frac{\Delta}{\Sigma_0} \, \frac{\pp/R_y }{Q_p} ,
\end{equation}
where $\Delta$ was defined in \eqref{Deltadef}. This shows that for the dilaton the Buscher rule $\Phi \to \Phi - \frac{1}{2}\log g_{yy}$ is once again equivalent to the simultaneous replacements $Q_1 \leftrightarrow Q_p$ and $\k R_y \leftrightarrow \pp /R_y$ since the constant prefactor in front of $\Sigma_0$ is invariant by itself.

\subsection{JMaRT uniqueness}

Having rewritten the NS5-decoupled JMaRT solutions in the form given in Eq.\;~\eqref{FinalMetricandBfield}, we observed that these supergravity fields are exactly those we obtained from the null-gauge construction studied in the previous sections when inserting the parametrisation \eqref{integersCFT1odd}--\eqref{L32def} for the $l_i,r_i$ gauging parameters in the general solutions in Eq.~\eqref{eq:metric-B-gen}.

We have thus shown that we are able to reproduce the full family of supergravity backgrounds collectively denoted as JMaRT. We now argue that these solutions exhaust the full set of consistent null-gauged models considered in this paper, by scrutinising the allowed ranges of the parameters and identifying physically equivalent solutions.

At first sight, by looking at the  metric, $B$-field and dilaton given in Eqs.~\eqref{FinalMetricandBfield} and \eqref{FinalDilaton}, one would expect that $\k,\m,\n$ and $\pp$ could take any integer value. Additionally, in writing these expressions we have fixed an extra degree of freedom by choosing a positive sign for $\Delta$. However, this parameter space is constrained. In the general case, the parameter $\pp$ is fixed in terms of the other three as in Eq.~\eqref{pkequalsn5mn}, though we will see below that in the limit in which the fundamental string charge $Q_1$ vanishes, its value becomes arbitrary. On the other hand, choosing the opposite sign for $\Delta$ simply results in an equivalent time-reversed configuration. This leaves us with arbitrary $\m,\n$ and $\k$ (for negative $\k$, the orbifold parameter is identified with its absolute value). Changing the sign of $\m,\n$ or $\k$ can be compensated by choosing the orientation of the circle coordinate $y$ or the $\SSS^3$ angles $\phi$ and $\psi$, respectively. Therefore we can take all of $\m,\n,\k$ to be non-negative. In addition, due to the $\m \leftrightarrow \n$ symmetry described below \eqref{FinalMetricandBfield}, we are free to restrict to $\m \geq \n$. 
Finally, based on spectral flow considerations we have argued in Section~\ref{sec:CFTanalysis} that we must restrict to angular momenta such that $\m \pm \n$ are odd, see the discussion below Eq.~\eqref{integersCFT1}. This clarifies the discussion about spin structures in \cite{Jejjala:2005yu}.
This excludes $\m=\n$ and so we conclude that the set of inequivalent configurations is given by 
\begin{equation}
    \k \, \geq 0 \ , \ \m > \n \geq 0,  \qquad  \m \pm \n \in 2 \mathbb{Z}+1, 
\end{equation}
which is precisely the principal range of values considered in \cite{Jejjala:2005yu}. As we shall discuss below, for $\k=0$ we take the limit such that one of the angular momenta vanishes and $\pp$ is generically kept non-zero and finite.

\subsection{Two-charge limits and novel non-BPS NS5-P solutions}

The expressions \eqref{FinalMetricandBfield} and \eqref{FinalDilaton} we have obtained for the metric, $B$-field and dilaton generated in the classical limit of the null-gauged models match exactly the solutions given in~\cite{Jejjala:2005yu}, which have $\k > 0$ (and where $\pp= n_5 \m \n/\k$). However, and as discussed above, they are presented in a form that is slightly more general and can be used to access somewhat delicate limits. In particular, we now examine two-charge limits.

There are two such limits that we can access. The first of these corresponds to NS5-F1 solutions, obtained by setting $Q_p=0$, which were analysed in~\cite{Jejjala:2005yu}. As shown by the identity \eqref{kn1pnp}, in order to keep $Q_1$ finite and arbitrary we need to do this carefully. More precisely, we also need to take the limit $\pp \to 0$ in such a way that the ratio $\pp/(R_y Q_p) = \k R_y /Q_1$ is finite. An analogous conclusion for taking $\n \to 0$ with $\pp/\n = n_5 \m/\k$ fixed is obtained by considering \eqref{pkequalsn5mn}. 

On the other hand, we can now similarly access a different limit leading to novel non-BPS NS5-P configurations. In this case, we take $Q_1=\k=\n=0$ while keeping the ratios $\k R_y/Q_1=\pp/(R_y Q_p)$ and $\k/\n = n_5 \m/\pp$ fixed. This allows $Q_p$ to take arbitrary values as needed. To the best of our knowledge, the metric and $B$-field for the non-BPS NS5-P solutions have not been presented in the literature. They take the following form:   
\begin{align}
\label{MetricandBnewNS5P}
    ds^2  \;=\; & \frac{1}{\Sigma_0}\Bigg[  - \left(\sinh^2\rhoo + \m^2 (\cos^2 \theta-1) + 1 - \frac{\pp^2}{n_5 R_y^2}  \right)   dt^2 - 2\, \frac{\pp }{n_5 R_y} \, \Delta\; dt dy \\
     &+\left(\sinh^2\rhoo + \m^2\cos^2 \theta  + \frac{\pp^2}{n_5 R_y^2}  \right)   dy^2  +  \, 2 \m \left( \Delta \, dt  - \frac{\pp}{R_y}  \, dy \right)\,  \sin^2 \theta \, d\phi  \nonumber\\
     & + n_5 \left( \sinh^2 \rhoo + \m^2 \right) \sin^2 \theta \, d\phi^2 +  n_5 \sinh^2 \rhoo  \, \cos^2 \theta \, d\psi^2 \Bigg] + n_5(d\theta^2 + d\rho^2)  \,, \nonumber \\[3mm]
 B \;=\;& \frac{n_5}{4\Sigma_0} \Big[\m^2 -1 +  \cosh(2\rho)   \Big]\cos(2 \theta) \, d\phi \wedge d\psi  + \frac{\m}{\Sigma_0} \left( \Delta \, dt  - \m \frac{\pp}{R_y} \, dy \right) \wedge  \cos^2 \theta \, d\psi \,, \nn 
\end{align}
with the dilaton given by the second expression in \eqref{FinalDilaton}, 
and where we now have
\begin{align}
\label{defsnewNS5P}
    \Sigma_0 \,=\, 
    \sinh^2\rho + \m^2 \cos^2\theta \ , \qquad 
    \Delta \,=\, \sqrt{n_5(\m^2-1) + \frac{\pp^2}{R_y^2}} \;.
\end{align}
%
Recall that, as discussed around Eq.~\eqref{eq:m-orb}, these solutions involve a fivebrane source at $\rho=0,\theta=\pi/2$ and a $\mathbb{Z}_{|m|}$ orbifold singularity at $\rho=0,\theta\neq \pi/2$.

Finally, we can restrict to the BPS cases by setting $\m=1$, as indicated by the supersymmetry conditions \eqref{susyrestricparam} and \eqref{susyrestricparamR}.  
Thus, in this limit we find $\Delta = \k R_y$ for the BPS NS5-F1 configuration, and $\Delta = \pp/R_y$ for the NS5-P one. In the former case, the first expression for the dilaton in  \eqref{FinalDilaton} then gives 
\begin{equation}
    e^{2\Phi}|_{\mathrm{NS5-F1}} \,=\, \frac{1}{Q_1} \, \frac{ \k^2 R_y^2}{\sinh^2\rho + \cos^2\theta + \k^2 R_y^2/n_5} \, ,
\end{equation}
which coincides with that of \cite{Martinec:2017ztd}, Eq.~(4.13), while for the latter the alternative expression in \eqref{FinalDilaton} yields
\begin{equation}
    e^{2\Phi}|_{\mathrm{NS5-P}} \,=\, \frac{1}{Q_p} \, \frac{(\pp/R_y)^2 }{\sinh^2 \rho + \cos^2 \theta} \,,
\end{equation}
which coincides with that of \cite{Martinec:2017ztd}, Eq.~(4.2). One can check that in both cases the metric and $B$-field match as well.

\subsection{$AdS_3$ limit and holography}

The $AdS$ limit of the geometries under consideration is obtained by taking the large $R_y$ limit, while keeping the charge $Q_1$ fixed. This describes the region of small radial distances (as compared with $Q_1$ and $R_y$). The energy and momenta $E R_y$ and $P_y R_y$ also stay fixed, such that the coordinates 
\begin{equation}
    \tilde{t} = t/R_y \ , \ 
    \tilde{y} = y/R_y
\end{equation}
are better suited for this region. The six-dimensional metric \eqref{FinalMetricandBfield} then takes the form of an orbifolded $AdS_3 \times \SSS^3$, namely
\vspace{-2mm}
\begin{align}\label{eq:ads-JMaRT-met}
\begin{aligned}
\!\!\!\!
ds^2 \;= ~ & 
	n_5 \left[ -\frac{1}{\k^2}\cosh^2 \rho \,  d\tilde{t}^2 +\frac{1}{\k^2}  \sinh^2 \rho \, d\tilde{y}^2  +  d\rho^2 
	\phantom{\left(-\frac{n}{\k}\right)^2 }	\right. 
	\cr
	&~~~~ {} \left. {} + d\theta^2+ \sin^2 \theta \left( d\phi 
	-\frac{\n}{\k} d\tilde{t}
	+\frac{\m}{\k} d\tilde{y}
	 \right)^2 
	%
	+ \cos^2 \theta \left( d\psi
	+\frac{\m}{\k} d\tilde{t}
	-\frac{\n}{\k} d\tilde{y}
	 \right)^2 \right] .\!\!\!\!\!\!\!
\end{aligned}
\end{align}
The orbifold singularity structure near $\tilde{y}=0$ depends on the common divisors between $\m,\n,\k$ and is described in~\cite{Jejjala:2005yu,Chakrabarty:2015foa,Martinec:2018nco}.
By means of the large gauge transformation
\be
\label{LGT}
\tilde\psi \;=\;
\psi
	+\frac{\m}{\k} \:\! \tilde{t}
	-\frac{\n}{\k} \:\! \tilde{y}  \,, 
	\qquad 
\tilde\phi \;=\;
\phi 
	-\frac{\n}{\k} \:\! \tilde{t}
	+\frac{\m}{\k} \:\! \tilde{y}\,,
\ee
one can formally re-absorb the contributions from the  angular momenta, that is, the terms depending on $\m$ and $\n$. 
This is related to the general holographic description of such configurations. They are interpreted as excited states in the  holographic symmetric orbifold CFT which can be constructed by considering $n_1n_5/\k$ identical strands of length $\k$ in their NS vacuum state and performing left-right asymmetric fractional spectral flow \cite{Chakrabarty:2015foa}.
The spectral flow charges are of the form 
\begin{equation}
\label{SFchargesHCFT}
    2\alpha = \frac{\m+\n}{\k} \ , \ 
    2\bar{\alpha} = \frac{\m-\n}{\k}
\end{equation}
which matches the intuition derived from \eqref{LGT} and provides yet another interpretation for the gauging parameters $l_2$ and $r_2$. 
Note that this is distinct from the \textit{worldsheet} spectral flow that was used in Section~\ref{sec:CFTanalysis}.
Interestingly, within this description the momentum per strand is given by $\m \n/\k$ and must be an integer number, which is a slightly more restrictive condition than the quantisation of $\pp$ discussed above. 

Furthermore, we also note that in this $AdS$ limit there seems to be no particular issues with the solutions with even $\m \pm \n$, see the discussion around Eq.\;\eqref{integersCFT1odd}. In particular, the case $\m=\n=0$ takes us back to the global $AdS_3$ vacuum. The present perspective shows that the cases that extend consistently to the full linear dilaton geometry, namely $\m \pm \n \, \in \, 2\mathbb{Z}+1$, belong to the RR sector (in the covering space) of the holographic CFT, while those that do not are characterised by having spectral flow charges \eqref{SFchargesHCFT} with even numerators, such that they correspond to NSNS states.

Moreover, in this $AdS_3$ limit the dilaton becomes constant, as can be seen from \eqref{FinalDilaton} since $\Delta \to \k R_y$ and $\Sigma_0 \to \k^2 R_y^2/n_5$. In other words, the rescaled harmonic function $\tilde{H}_1$ associated to the fundamental string charges approaches the constant value $Q_1$, see Eq.\;\eqref{Q1Qpb2first}. In terms of the actual harmonic function $H_1$, this roughly corresponds to the usual \textit{dropping the "1+"} term, as is usual in such decoupling limits~(see e.g.~\cite{Bena:2018bbd}). 

In terms of the null-gauged description, there is an intuitive way of understanding this $AdS_3 \times \SSS^3$ limit. Indeed, the upstairs model already contains an $SL(2,\R)\times SU(2)$ factor, complemented by the novel $\R_t \times \SSS_y^1$ factor. For large $R_y$, the gauging parameters associated to the former are $l_{1,2}\sim r_{1,2} \sim {\cal{O}}(1)$, while those corresponding to the latter grow parametrically large as $l_{3,4}\sim r_{3,4} \sim {\cal{O}}(R_y)$. Thus, we are mostly gauging away the extra directions $t$ and $y$.

\subsection{From $AdS$ back to the linear dilaton background}

At this point, it is interesting to go back to the intuition developed within the sigma model description presented at the beginning of Section~\ref{sec:nullgauging}. There, we argued that, after integrating out the gauge fields, the gauging procedure induces a deformation given by including an additional term of the form $\cJ \bar{\cJ}/\Sigma$ to the action, see Eq.~\eqref{eq:g-terms-int-out}. We can simplify the discussion by working in the $t=y=0$ gauge, such that the currents $\cJ$ and $\bar{\cJ}$ are nothing but linear combinations of the diagonal currents of the $SL(2,\R)$ and $SU(2)$ WZW models. In the $AdS$ limit, we have seen that the coefficient $1/\Sigma$ approaches a constant value, such that the induced contribution becomes $\cJ \bar{\cJ}$, giving a simple marginal deformation of the worldsheet theory.  

The situation is to be contrasted with that of \cite{Giveon:2017nie,Asrat:2017tzd,Giveon:2017myj}. There, the authors make use of the null-gauging formalism to introduce a $J \bar{J}$ deformation for the $AdS_3$ worldsheet theory, the crucial difference being that the current under consideration corresponds to the $J^-$ instead of $J^3$. For related work, see~\cite{Chakraborty:2020yka}. Based on \cite{Kutasov:1999xu}, this procedure is interpreted as the dual of the so-called single-trace $T\bar{T}$ irrelevant deformation of the holographic CFT. Such deformation triggers a controlled flow to the UV, which is realised in the dual geometry by effectively \textit{reinserting the "1+" term} in the harmonic function associated to the fundamental string sources. This produces an asymptotically linear dilaton geometry, i.e. an NS5-decoupled background, such that the above UV flow is thought of as leading to a realisation of little string theory. 

The parallel to our construction can thus be made more general. As described above, for large $R_y$ we know that  $\Sigma$ becomes constant, and our $\cJ \bar{\cJ}$ deformation on the worldsheet made up of $J^3$ and $K^3$ (together with their antiholomorphic counterparts) has a much less dramatic effect, as it produces a sort of large gauge transformation which, however, does not further modify the $AdS_3$ asymptotics, where the dilaton stays constant. On the other hand, when moving away from the $AdS$ limit by keeping $R_y$ finite one recovers the full non-trivial coordinate dependence of the function $\Sigma$, which sits at the denominator of various terms in the supergravity fields. This modifies the effect of the $\cJ \bar{\cJ}/\Sigma$, which now \textit{does} take us back to the full asymptotically linear dilaton background described by Eqs.~\eqref{FinalMetricandBfield} and \eqref{FinalDilaton}. This effect has also recently been observed in a larger class of solutions in~\cite{Martinec:2020gkv}.

\section{Discussion}
\label{sec:discussion}

In this paper we have analysed all consistent backgrounds within a general class of null-gauged WZW models. We showed that the (NS5-decoupled) JMaRT family, and limits thereof, are the unique supergravity backgrounds that arise in these models. We also showed that the metric and $B$-field can be written explicitly in terms of the integers $\k,\m,\n,n_5$ and the modulus $R_y$, while for the dilaton one needs further include the ratio $n_1/V_4$ (or $n_p/V_4$).

Our analysis makes the connection between the worldsheet and geometric descriptions quite explicit. 
In the supergravity solutions, imposing absence of CTCs implies $l_3=r_3$, which excludes horizons. The converse statements are also true: excluding horizons implies $l_3=r_3$, which excludes CTCs.
On the other hand, the condition $l_3=r_3$ is necessary but not sufficient for smoothness (up to orbifold singularities), which further requires the quantization of  $\k,\m,\n$.
This is consistent with the interpretation of the allowed configurations as a family of black hole microstates.

At the level of the worldsheet CFT, we have shown that a consistent spectrum is obtained if and only if the gauging parameters are given in terms of three independent integers $\k,\m,\n$ together with $n_5$ and $R_y$, as in \eqref{integersCFT1odd}--\eqref{pkequalsn5mn}. We explicitly rewrote the JMaRT metric and $B$-field in terms of these quantities, which enabled us to  completely bypass the usual, somewhat cumbersome, supergravity parametrisation in Eqs.~\eqref{eq:JM-orig}, \eqref{eq:JM-orig-B}.

Our parametrisation also provides a clear and important physical understanding of the quantity $\pp=n_5 \m \n/\k$. In the $AdS_3$ decoupling limit, the quantity $\m \n/\k$ has previously been interpreted as being the momentum per strand in the holographically dual symmetric product orbifold CFT~\cite{Chakrabarty:2015foa}. We have uncovered the direct role of $\pp$ in the (asymptotically linear dilaton) supergravity solutions, as being the quantity that is T-dual to $\k$, where the T-duality is performed along the $y$ circle. 

As we mentioned in the Introduction, one of the motivations of our systematic analysis was the possibility of finding new backgrounds. Although our uniqueness proof means that the set of models we analysed does not have more general backgrounds than the JMaRT family, we have exhibited a novel sub-family of two-charge non-BPS NS5-P backgrounds that arise from a non-trivial limit, see Eqs.~\eqref{MetricandBnewNS5P} and \eqref{defsnewNS5P}. We observed that 
in the core of the solutions but
away from the fivebrane source, the solutions involve a $\mathbb{Z}_{\m}$ orbifold singularity. To our knowledge, these solutions have not appeared before in the literature.

We expect that our results will be useful in analysing generalisations of the models studied here, either by changing the currents being gauged to include non-Cartan generators of the non-Abelian factors of the upstairs group, or by changing the upstairs group, or both. Our systematic approach should enable generalisations to be investigated in a similar way. For instance, there are multi-centre non-BPS generalizations of the JMaRT family~\cite{Bena:2015drs,Bena:2016dbw,Bossard:2017vii}.

Besides this, within the models considered in this work there remain several unanswered questions. For instance, since we have control over these theories exactly in $\alpha'$,   there are many interesting correlation functions that can be computed.  We intend to report an analysis of such correlators in the near future.

The results we have obtained, and the possibilities they open up for future work, offer the prospect of improving our understanding of little string theory and the corresponding non-AdS holography. Furthermore, it is tempting to wonder about extending some of these ideas beyond the fivebrane decoupling limit into the full asymptotically flat regime.

Having an exact worldsheet description of heavy pure states, far from the vacuum of the theory, is rare and valuable. Such models allow us to study aspects of black hole microstates that are smeared out in supergravity, and so cannot be studied with supergravity techniques. This offers the tantalising prospect of obtaining a quantitative understanding of the microscopic degrees of freedom of black holes.

\acknowledgments

We thank Emil Martinec,  Stefano Massai, Sami Rawash, Michele Santagata and Jerem\'ias Aguilera-Damia  for discussions.
The work of D.B.~was supported by the Royal Society Research Grant RGF\textbackslash R1\textbackslash181019.
The work of S.I.~was partially supported by the CONICET fellowships PICT-2016-1358  and PUE-IAFE22920160100060CO.
The work of N.K.~was supported by the Leverhulme Trust under grant no.~RPG-2018-153 and by the ERC Consolidator Grant 772408-Stringlandscape. 
The work of D.T.~was supported by a Royal Society Tata University Research Fellowship. 

\appendix


\bibliographystyle{JHEP}
\bibliography{refs}

\end{document}